\documentclass[a4paper,aps,pre,floatfix,twocolumn,showpacs,nofootinbib,superscriptaddress]{revtex4-1}
\usepackage{graphicx}
\usepackage{latexsym,amsmath,verbatim}
\usepackage{color}
\usepackage{hyperref}
\usepackage{multirow}

\newcommand{\be}{\begin{equation}}
\newcommand{\ee}{\end{equation}}

\usepackage[normalem]{ulem}

\begin{document}
%\begin{CJK*}{}{}

\title{General three-state model with biased population replacement: \\
Analytical solution and application to language dynamics}

\author{Francesca Colaiori}
\author{Claudio Castellano}
%\footnote{author to whom correspondence should be addressed: claudio.castellano@roma1.infn.it}}
\author{Christine F. Cuskley}
\affiliation{Istituto dei Sistemi Complessi (ISC-CNR), via dei Taurini 19, I-00185 Roma, Italy}
\affiliation{Dipartimento di Fisica, Sapienza Universit\`a di Roma, P.le A. Moro 2, I-00185 Roma, Italy}
\author{Vittorio Loreto}
\affiliation{Dipartimento di Fisica, Sapienza Universit\`a di Roma, P.le A. Moro 2, I-00185 Roma, Italy}
\affiliation{Istituto dei Sistemi Complessi (ISC-CNR), via dei Taurini 19, I-00185 Roma, Italy}
\affiliation{ISI Foundation, Via Alassio 11/C,  Torino, Italy}
\author{Martina Pugliese}
\affiliation{Dipartimento di Fisica, Sapienza Universit\`a di Roma, P.le A. Moro 2, I-00185 Roma, Italy}
\affiliation{Istituto dei Sistemi Complessi (ISC-CNR), via dei Taurini 19, I-00185 Roma, Italy}
\author{Francesca Tria}
\affiliation{ISI Foundation, Via Alassio 11/C,  Torino, Italy}

\date{\today}

\begin{abstract}
Empirical evidence shows that the rate of irregular usage of English
verbs exhibits discontinuity as a function of their frequency: the 
most frequent verbs tend to be totally irregular. We aim
to qualitatively understand the origin of this feature by studying
simple agent--based models of language dynamics, where each agent
adopts an inflectional state for a verb and may change it upon
interaction with other agents. At the same time, agents are replaced
at some rate by new agents adopting the regular form.  In models with only
two inflectional states (regular and irregular), we observe that
either all verbs regularize irrespective of their frequency, or a
continuous transition occurs between a low-frequency state where the
lemma becomes fully regular, and a high frequency one where both
forms coexist. Introducing a third (mixed) state, wherein agents may
use either form, we find that a third, qualitatively different behavior
may emerge, namely, a discontinuous transition in frequency. We introduce
and solve analytically a very general class of three--state models
that allows us to fully understand these behaviors in a unified framework. 
Realistic sets of interaction rules, including the well-known Naming 
Game (NG) model, result in a discontinuous transition, in agreement 
with recent empirical findings. We also point out that the distinction 
between speaker and hearer in the interaction has no effect on the 
collective behavior. The results for the general three--state model, 
although discussed in terms of language dynamics, 
are widely applicable.
\end{abstract}
  
%\pacs{05.40.Fb, 89.75.Hc, 89.75.-k}
\maketitle
%\end{CJK*}

\section{Introduction}

Language is structured by rules~\cite{general1, general2} - but 
linguistic rules often have exceptions. 
This fact kindled a long-standing debate in cognitive science
centering on how individual learners accommodate rule sets rife with
exceptions (e.g., see Refs.~\cite{pinker_words, mcclelland_words}). 
However, how exceptions
arise and evolve over time within a language system remains a largely
open question.

\begin{figure}[h!]
\includegraphics*[width=\columnwidth]{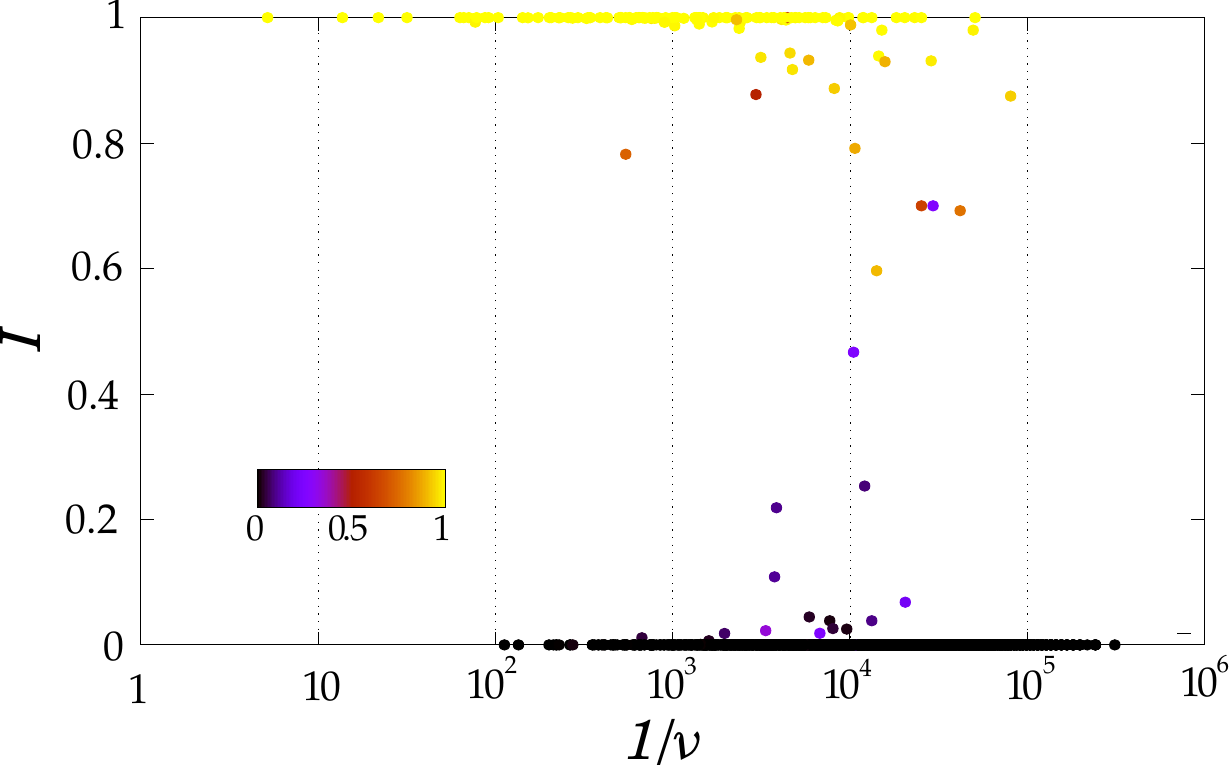}
\caption{(color online)
Plot of $I$, the fraction of times a verb is used in the irregular past tense 
form as a function of the inverse of its frequency of usage $\nu$, 
for 2840 english verbs in the  CoHA corpus for the decade 1980-1989. 
The color (shade) of each symbol represents the average value of $I$ across 16 
decades between 1830 and 1989. (Adapted from Ref.~\cite{cuskley_2014}).}
\label{figplos}
\end{figure}

The study of the English past tense is
widely used as an exemplar of the interplay between rules and
exceptions~\cite{pinker_pasttense, rules_exc1, rules_exc2, rules_exc3}.
A recent study of historical corpus data~\cite{cuskley_2014} looks at rules
in the language system rather than individual learners, shedding light 
on the relationship between the verb frequency and regularity 
(Fig.~\ref{figplos}).
Each verb in the language can be
characterized by $I$, the fraction of irregular past tense tokens
over the total number of tokens in the past tense, and $\nu$, its 
frequency of usage. An interesting transition is found in the behavior
of $I$ as a function of $\nu$: regular verbs dominate
the low frequency range, while most irregular verbs are located at
higher frequencies (see also Refs.~\cite{lieberman,german}). 
For intermediate values of $\nu$ fully regular ($I=0$) and fully irregular 
($I=1$) verbs coexist.
Only a small subset of verbs exhibit both regular and irregular forms
($0<I<1$),
and occur primarily in a rather narrow range of frequencies between the
dominant regular and irregular states. 

The work presented in this paper takes a theoretical approach to the
relationship between rules and exceptions in a population of
interacting speakers.  We investigate the dynamics of a set of very
simple agent-based models aimed at describing the fundamental
mechanisms by which rules and exceptions may be shared or disappear in
a population, in the same spirit of the Naming Game
(NG)~\cite{NG_depth, stat_lang} investigations of the emergence of
shared naming conventions.  We consider a single lemma and examine
two-state and three-state models.  In a two-state model, an individual
has an internal inflectional inventory which can contain either the
regular (R) or irregular (I) inflection for the lemma.  In a
three-state model, the inflectional state can be either R, I or mixed
(M). The mixed state represents intra-speaker
variation~\cite{labov_soc}, where an individual may accommodate both
regular and irregular forms for a single word~\cite{diversity}.  For
example, there is evidence that regular and irregular past tense verb
forms are simultaneously known and potentially used in seemingly free
variation within a single speaker (e.g., both~\textit{sneaked}
and~\textit{snuck} are acceptable~\cite{diversity}).  Agents are
endowed at the start of the dynamics with some inflectional state for
the word and engage at rate $\nu$ in pairwise interactions, i.e.
one of them utters the verb under consideration and the other listens to it.  
During interaction, the inflectional states of the speaker and hearer change
according to a pre-defined set of interaction rules.  In addition to
interaction events, we implement a \textit{replacement} mechanism: at
rate $r$ an agent is replaced with a ``child'' who engages in
\textit{overregularization}: these ``child'' agents assume the regular
inflection applies to all words in the vocabulary, representing a
known bias of child learners, \cite{berman_1981, bybee_children_1982,
  dense_corpus}.  Replacement represents turnover in the population:
as a child learner enters, an adult leaves (i.e., is replaced), so
that the population size remains constant.

We first approach some specific models analytically, within the framework
of mean-field theory. This analytical approach allows for quantitative
predictions regarding which state a population of speakers will reach
given a particular set of interaction rules and the type of transition
which may occur depending on the ratio between the frequency and replacement
rates.
It turns out that different interaction rules lead to qualitatively different
types of behavior. Two-state models lead to total regularization or to
a continuous transition as a function of $r/\nu$. 
Three-state models, on the other hand, have the potential to
exhibit a discontinuous transition with some highly
frequent, mostly irregular words, reminiscent of findings in empirical
data. To understand these variations and understand their origin,
we introduce a very general three-state model, encompassing all
possible sets of interaction rules that do not favour either the
regular or irregular state in their outcome. We provide a formal
solution for any value of the model parameters and study in detail
the conditions under which no transition (i.e., total regularization)
occurs, and the conditions under which we observe a continuous or
discontinuous transition.
Our analysis also shows that assuming
asymmetric influence of the speaker over the hearer in the interaction
has no effect on the collective behavior. The results for this general
model are discussed in terms of language dynamics, but their
applicability is fully general: they give the solution for any
three-state model of population dynamics with unbiased interaction
rules and biased replacement.

\section{Two-state models}

Let us start by considering two-state models, where each individual
can be found in one of two possible rule states, regular (R) or
irregular (I). While in this case the two states represent regular and
irregular inflections, this framework can represent any set of binary
options (e.g., the choice between two possible words to name an
object, two alternative opinions on a given topic, etc.). As a
specific example, the Abrams-Strogatz
model~\cite{Abrams-Strogatz_2003} used a two--state approach to
examine the dynamics of endangered languages, providing a prime
example of the dynamics of languages in competition more generally.

Using this general two--state model approach, we focus here on the
binary options of regular (R) or irregular (I) inflection for a single
word, characterized by its frequency of usage, $\nu$.  At each
interaction step two agents are selected at random (i.e., mixing is
homogeneous) and assigned the role of either speaker or hearer. With
probability $\nu$ they engage in a pairwise interaction
(i.e. the speaker utters the verb under consideration to the hearer), 
which affects their inflectional states according to specific 
interaction rules.
Then, with probability $r$ one individual in the population is
replaced with a ``child'' having its inflectional state set to R.
This part of the dynamics mimics the turnover of some segment of the
adult population into child learners at rate $r$, keeping the
population size $N$ fixed, and assuming over-regularization behavior
in new learners. In this way the population replacement is biased
towards one of the two options, in this case, regularity.  The
quantity $1/r$ can be interpreted as the life expectancy of an
individual in the population.

Among the possible two--state interaction rules, we consider the sets
presented in Table~\ref{2stateTable}.
\begin{table*}
\begin{tabular}{lllllllllll}
\hline
\hline
\multicolumn{2}{l}{\bf {Before}}&\phantom{abcdefgh}& \multicolumn{8}{c}{\bf After}\\
\hline\hline
\multicolumn{2}{l}{} &&\multicolumn{2}{l}{Model A:}  &\phantom{aa}& \multicolumn{2}{l}{Model B:}&\phantom{aa}& \multicolumn{2}{l}{Model C:}\\
\multicolumn{2}{l}{} &&\multicolumn{2}{l}{Irregular-biased}  &\phantom{aa}& \multicolumn{2}{l}{Regular-biased}&\phantom{aa}& \multicolumn{2}{l}{Speaker leads}\\
\hline
{Speaker}& {Hearer} &&Speaker &Hearer&& Speaker &Hearer &&Speaker &Hearer\\
\hline
\hline
R & R && R & R && R & R && R & R\\
R & I  &&  I & I  && R & R && R & R\\
I &  R && I  & I  && R & R && I & I\\
I &  I && I  &  I  && I  & I && I &I\\
\hline
\hline 
\end{tabular}
\caption{Interaction rules for the two-state models.  The two columns
  on the left refer to the status of speaker and hearer prior to
  interaction. The next three pairs of columns refer to the status of
  speaker and hearer after the interaction in the three models. }
\label{2stateTable}
\end{table*}
In the {\it Irregular--biased} (A) and {\it Regular--biased} (B) models, 
the speaker and the hearer roles are symmetric; 
in other words, which agent identifies
as speaker or hearer is irrelevant, but the presence of an I(R) state
in the interaction leads the rules. In A(B) an agent switches to the
I(R) state whenever interacting with a partner in the I(R) state,
regardless of which agent is the speaker and which the hearer. In
these cases the speaker can affect the hearer's state, and the hearer
can also affect the speaker's. In the {\it Speaker leads} model (C) 
the roles are not symmetric: the speaker never changes its state
and the hearer always adopts the state of the speaker.

The {\it Irregular--biased} model  is perfectly equivalent to one of the most fundamental models
of non--equilibrium statistical physics: the contact
process~\cite{Marrobook_1999}.
The temporal behavior of this model is easily understood by
writing down the mean-field evolution equation for the density
$\rho_I$ of individuals in the I state (the density of R individuals,
$\rho_R$, being trivially $1-\rho_I$)
\be
\dot{\rho}_I = - r \rho_I + 2 \nu \rho_I (1-\rho_I).
\label{2state}
\ee
Equation~(\ref{2state}) is solved straightforwardly and yields, for
any initial configuration $\rho_I(0), \rho_R(0)=1-\rho_I(0)$
\be
\rho_I(t) = \frac{2/n-1}{2/n+\left(\frac{2/ n-1}{\rho_I(0)}-2/ n
\right) e^{-r(2/n-1)t}}
\label{caseA}
\ee
where $n=r/\nu$.

For long time scales, the system reaches (for any initial configuration)
a stationary state which
exhibits a continuous transition for a critical value $n_c=2$,
between a fully regular state ($\rho_I=0$) for $n>n_c$, and a state 
with individuals in both the R and the I state ($\rho_I>0$):
\be
  \rho_I =
  \left\{ 
    \begin{array}{lr}
      0                \;\; & \; n \geq n_c  \\  
      1- \frac{n}{2}  \;\; & 0 \leq n < n_c.
    \end{array} 
  \right. 
  \label{caseAstationary}
\ee

The solutions for the {\it Regular--biased} and {\it Speaker leads} 
models are obtained from Eq.~(\ref{caseA})
by simply replacing $n \rightarrow -n$ and $n \rightarrow \infty$,
respectively, and both result in an exponential relaxation to the
stationary fully regular state ($\rho_I=0$) for any physical value of
$n$.  Unlike in case A, in these two cases the interaction rules are
biased in favor of R (case B) or unbiased (C) and they cannot
compensate for the increase in R states due to replacement,
leading to a fully regular absorbing state. 

As it will be demonstrated in Sec.~\ref{General2}, 
no two-state model can give rise to
a discontinuous transition between the fully regular state and a
state with $\rho_I>0$.  
Empirical data, however, exhibit such a discontinuous transition, and
research shows that speakers can accommodate regular and irregular
forms simultaneously \cite{diversity}. For these reasons, we now turn
our attention to a more complex modeling scheme that integrates a
third, mixed state (M), wherein agents accommodate either the R or I.
We will show that introducing this, psychologically plausible, mixed state,
a qualitatively different behavior appears, namely, a discontinuous
transition in regularity, reminiscent of empirical data.

\section{Three-state models}

In three--state models there are still only two alternative
inflections that can be applied to a word (R and I) during an
interaction event, but internally, each
individual can be in one of the three possible states: R (regular), I
(irregular) and M (mixed). In the mixed state the individual can
accommodate both R and I forms; this accounts for agents undecided on
which is the correct form to use, or that consider both the regular
and the irregular form acceptable.

The study of three-state models has a long history in the
investigation of language dynamics (for a review
see~\cite{Castellano_RMP_2009}). In particular, Wang and Minett
proposed~\cite{Wang_Minett_2005_TRENDS_Ecology,Minett_Wang_2008}
deterministic models for the competition of two languages, that
included a third potential state of bilingual individuals. Castell\'o
et al.~\cite{Castello_2006} proposed a modified version of the voter
model~\cite{clifford_1973,holley_1975} to examine language which
included bilingual individuals, the so-called AB model. Synonymy, the
possibility for having multiple potential names for a single meaning
(much like multiple inflections for a single verb as in the mixed
state) has also been examined in the classic Naming Game (NG)
model~\cite{NG_original_2006,NG_depth}.
The Naming Game and its variants have examined structures of
increasing complexity, often including agents who can have multiple
internal states, from the categorisation of colors~\cite{puglisi_2008,mukherjee2011aging}
to basic syntactic structures~\cite{tria_2012}.

We now study the dynamics of three specific
examples in the class of three-state models, with different
microscopic rules leading to qualitatively different behaviors (see
Table~\ref{3stateRules}). The first set of
rules is known as the Naming Game. 

\begin{table*}
\begin{tabular}{lllllllllll}
\hline
\hline
\multicolumn{2}{l}{\bf {Before}}&\phantom{abcdefgh}& \multicolumn{8}{c}{\bf After}\\
\hline\hline
\multicolumn{2}{l}{} &&\multicolumn{2}{l}{Model NG}  &\phantom{aaaa}& \multicolumn{2}{l}{Model CT}&\phantom{aaaa}& \multicolumn{2}{l}{Model NT}\\
\hline
{Speaker}& {Hearer} &&Speaker &Hearer&& Speaker &Hearer &&Speaker &Hearer\\
\hline
\hline
R & R &&  R & R  &&  R & R  &&  R & R \\
\hline
R & I  &&  R & M  && R & M && M & M \\
\hline
R & M && R  &R   && R &M && R  & R \\
   &    &&  &     &&       &  &&  M & M \\
\hline
I & R   && I & M &&   I  & M && M & M \\
\hline
I & I   &&   I & I     &&  I & I   &&  I & I \\
\hline
I & M && I  &I   && I &M && I  & I \\
   &    &&  &     &&       &  &&  M & M \\
\hline
M & R && M  &M   && R & R && R  & R \\
   &    && R&R     &&      I & M &&  M & M \\
\hline
M & I  &&  I & I  &&  I & I && I & I  \\
   &    &&  M & M &&     R & M && M & M \\
\hline
M & M && I & I & & R & M &&  I& I \\
   &     &&    R & R  && I &M  &&   R & R \\
      &     &&    &   &&  &  &&   M & M \\
\hline
\hline
\end{tabular}
\caption{Interaction rules for the three examples of three-state
  models. The two columns on the left refer to the status of speaker
  and hearer prior to the interaction. The following three pairs of
  columns refer to the status of speaker and hearer after the
  interaction for the three different rule sets:
the Naming Game (NG), the Continuous Transition (CT) model and the 
No Transition (NT) model. When two alternative
  inflections are possible because of a mixed state, the probability
  of each of them is $1/2$. In the only case with three alternative outcomes
  (I,I) and (R,R) occur each with probability 1/4, while (M,M) occurs probability 1/2. }
\label{3stateRules}
\end{table*}

\subsection{The Naming Game with biased replacement: A three--state
  model with a discontinuous transition}\label{sec:NGBR}

The interaction dynamics of the Naming Game with three states are as
follows: first, at each time step a speaker and a hearer are selected
at random. With the probability $\nu$ they interact; 
the speaker conveys to the hearer either the R or I form
depending on his inventory (if in the mixed state he utters R or I
with equal probability). If the hearer's inventory contains the
inflection used in the utterance, both agents update their inventories
keeping only the form involved in the interaction. Otherwise, the
hearer adds the form to his inventory (thus switching to the mixed
state). Table~\ref{3stateRules} (first four columns) summarizes these
interaction rules. In addition to these rules, the population turnover
is implemented as in the previous two-state models: at each time step 
an individual is selected at random and, with probability $r$, is replaced 
by a new individual in state R.

In a generic three-state model, two densities are needed to specify
the global state of the system. We choose $\rho_I$ and $\rho_R$, the
density of the mixed state being $\rho_M=1-\rho_R-\rho_I$. The
mean-field equations for the Naming Game with biased replacement are:
\be
\left\{
\begin{array}{lll}
\dot{\rho}_I &=& - r \rho_I + \nu (1+\rho_R^2 - 2\rho_R-\rho_I) \\
\dot{\rho}_R &=&  r (1-\rho_R) + \nu (1+\rho_I^2 - 2\rho_I-\rho_R).
\label{MF_NGWR}
\end{array}
\right.
\ee
Notice that the equations for the usual Naming Game with three
states~\cite{baronchelli_2007,NG_depth} are recovered by setting $r=0$
and $\nu=1$.
\begin{figure}[h!]
\includegraphics*[width=\columnwidth,height=5cm]{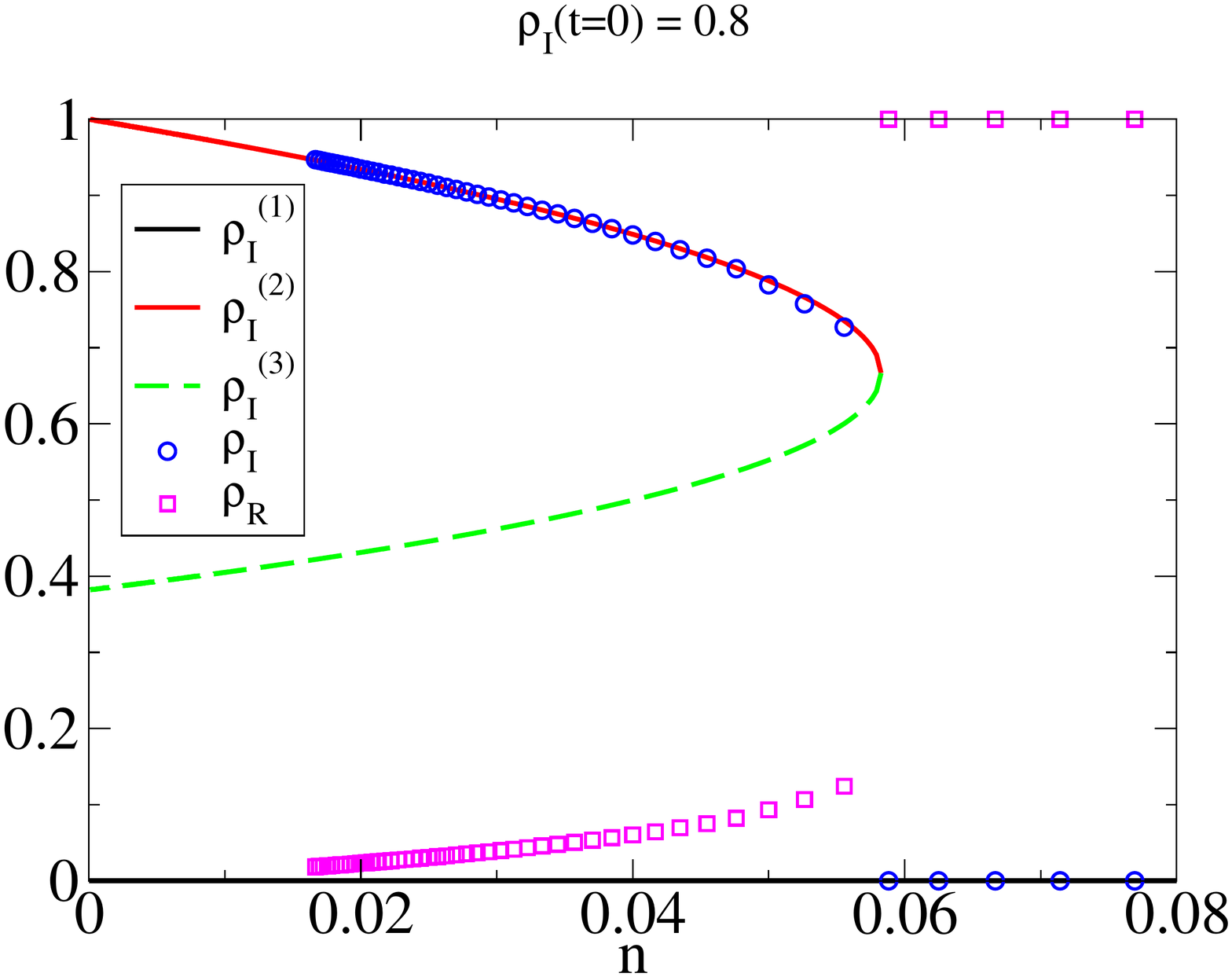}\\
\vspace{-0.5cm}
\includegraphics*[width=\columnwidth,height=5cm]{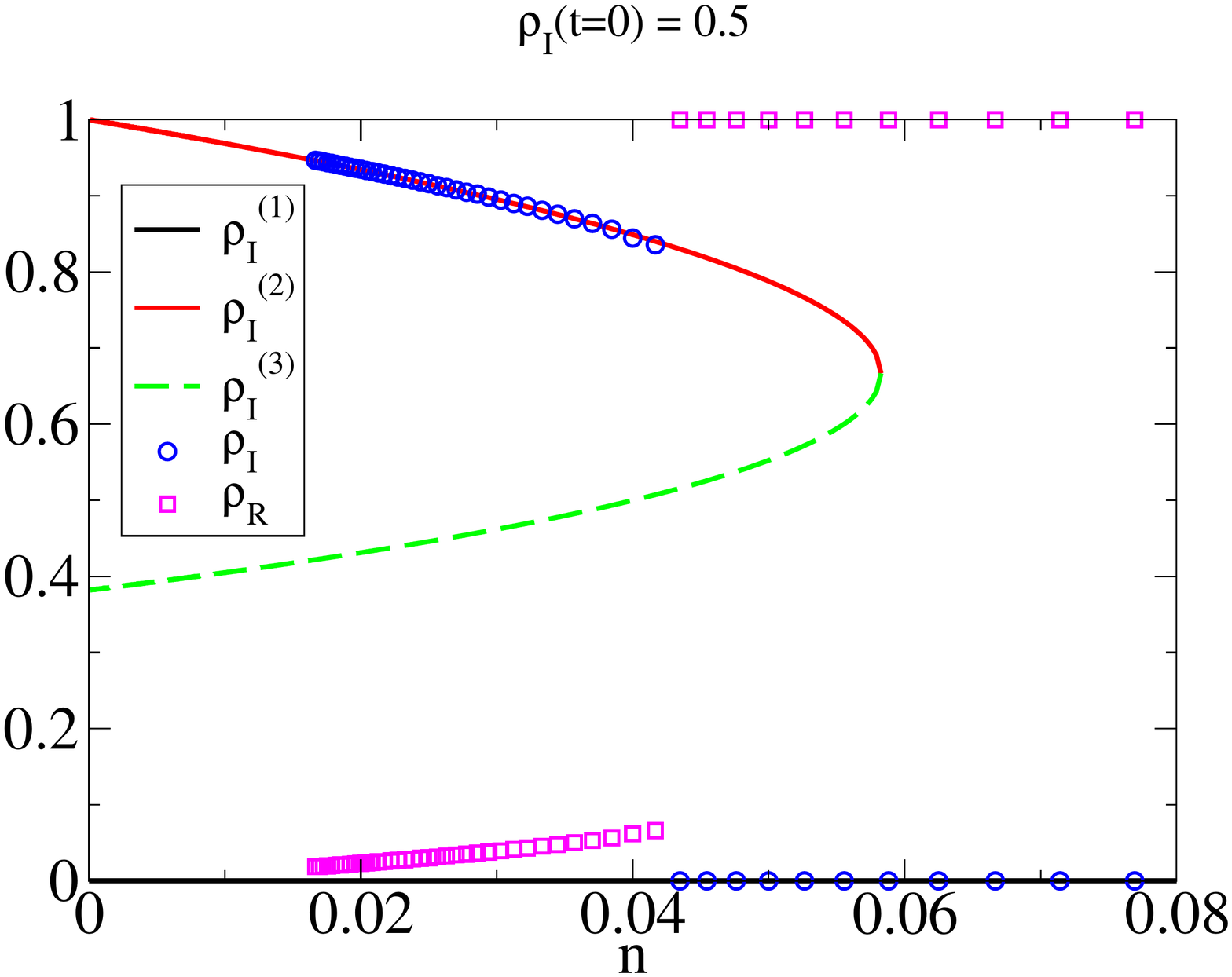}\\
\vspace{-0.5cm}
\includegraphics*[width=\columnwidth,height=5cm]{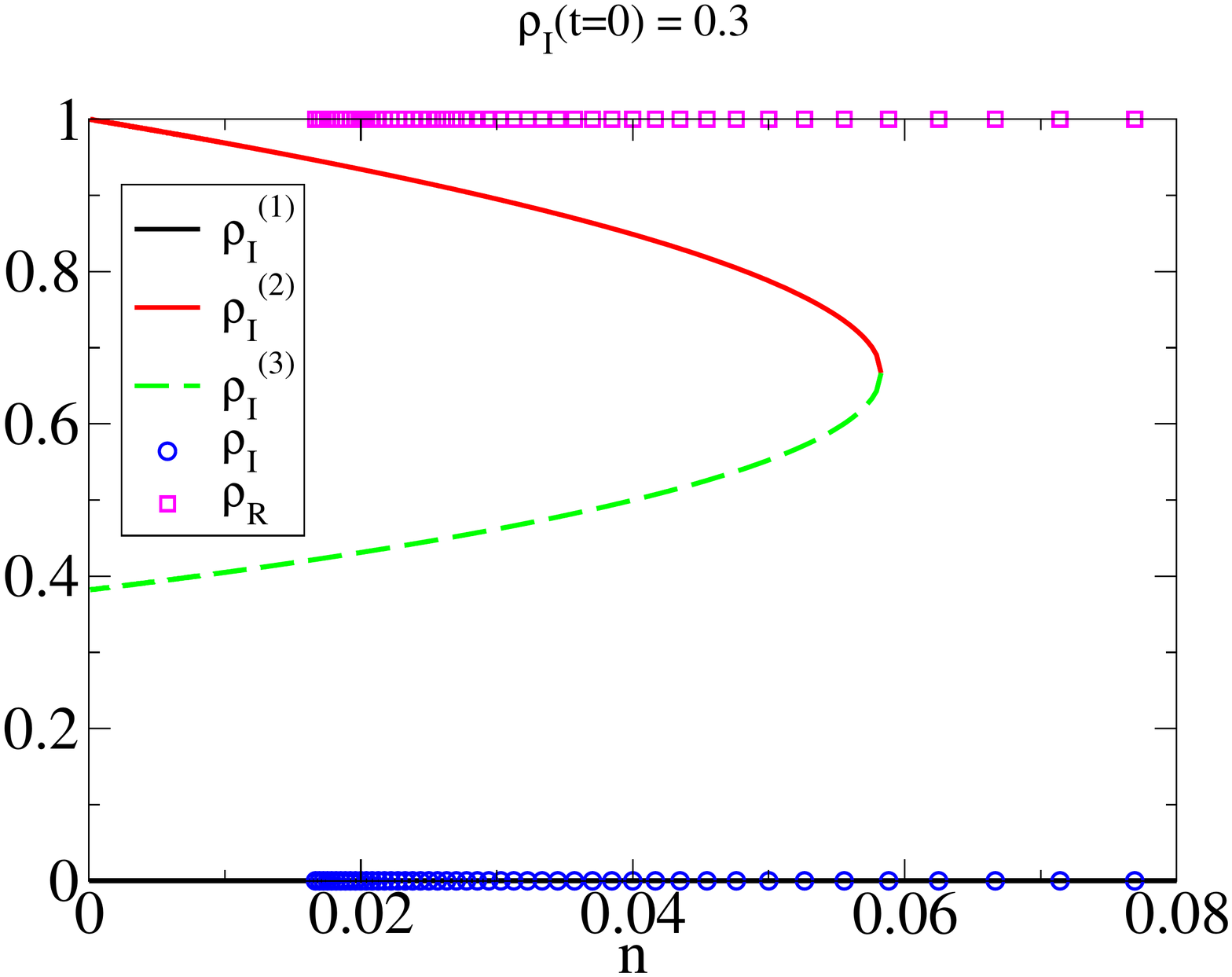}
\caption{(color online) Phase diagram of the NG model as a function of 
$n=r/\nu$.
The black line indicates the stationary solution $\rho_I^{(1)}$, 
while green (light gray) and red (dark gray) curves indicate 
solutions $\rho_I^{(2)}$ (stable)  and $\rho_I^{(3)}$ (unstable), respectively.  
Symbols are the stationary values of the fraction of irregulars $\rho_I$ 
and regulars $\rho_R$ from numerical simulations with
different initial conditions: from top to bottom $\rho_I(t=0)=0.8,0.5$ and 
$0.3$. The three panels show the dependence of the stationary state on 
the initial condition.}
\label{stationary_solutions}
\end{figure}
Imposing the stationarity condition, after some algebra one finds that
the density of individuals in the irregular state is given by the
fourth order equation
\be
\rho_I \left(1 + n \right)^3 = \rho_I^2 (\rho_I-2)^2,
\ee
where $n=r/\nu$. One solution is, for any $n$, the trivial value
$(\rho_I^{(1)},\rho_R^{(1)})=(0,1)$, corresponding to the fully
regular state. Regarding the three other solutions, since $(1+n)^3$ is
always larger than 1, it follows that one solution ($\rho_I^{(4)},
\rho_R^{(4)}$) is always real but unphysical (being larger than 1),
while the two others are complex for $n > {2\sqrt[3]{4}/3-1)} \approx
0.0583$. Below this critical value (corresponding to a saddle-node 
bifurcation), these two solutions are real and physical:
{\small{
\begin{eqnarray}
\rho_I^{(2)} &=& \frac{4}{3} \left[1 + \cos 
\left(\frac{\cos^{-1}\left[\frac{27}{16} 
\left(1+n\right)^3-1\right]}{3} + 
\frac{4\pi}{3}\right) \right]\\
\rho_I^{(3)} &=& \frac{4}{3} \left[1 + \cos 
\left(\frac{\cos^{-1}\left[\frac{27}{16} 
\left(1+n\right)^3-1\right]}{3} + 
\frac{2\pi}{3}\right) \right]
\end{eqnarray}
}}
with the stationary value of $\rho_R$ given by
\be
\rho_R^* = 1 - \sqrt{\left(1+n\right) \rho_I^*} \ .
\label{3stateVittorio_rhoR}
\ee
For $n=0$ the solutions converge to the values found for the usual
Naming Game~\cite{baronchelli_2007,NG_depth}: 
$(\rho_I^{(2)},\rho_R^{(2)})=(1,0)$, and 
$(\rho_I^{(3},\rho_R^{(3)})=((3-\sqrt{5})/2,(3-\sqrt{5})/2)).$

\begin{figure}
\includegraphics*[width=0.5\columnwidth]{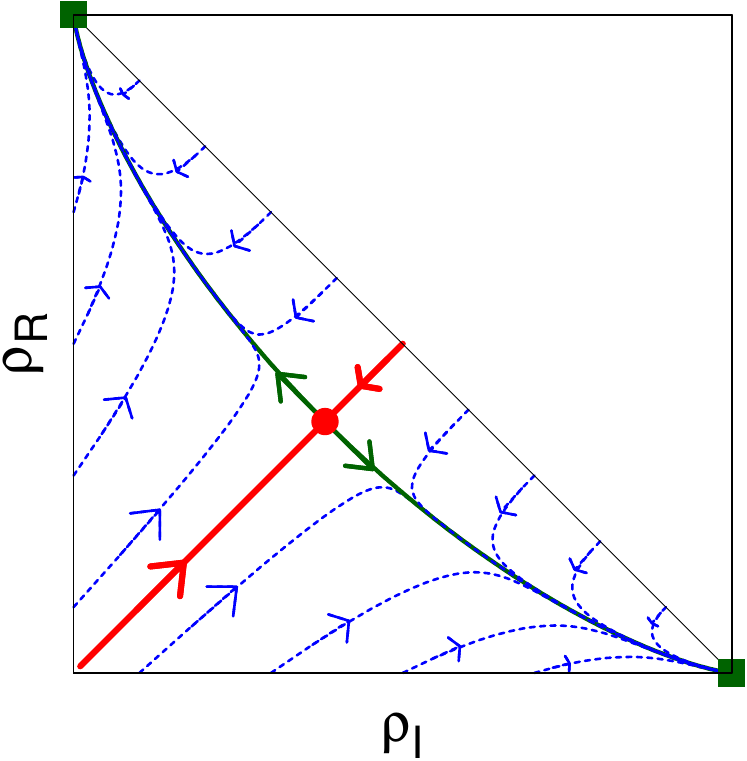}\includegraphics*[width=0.5\columnwidth]{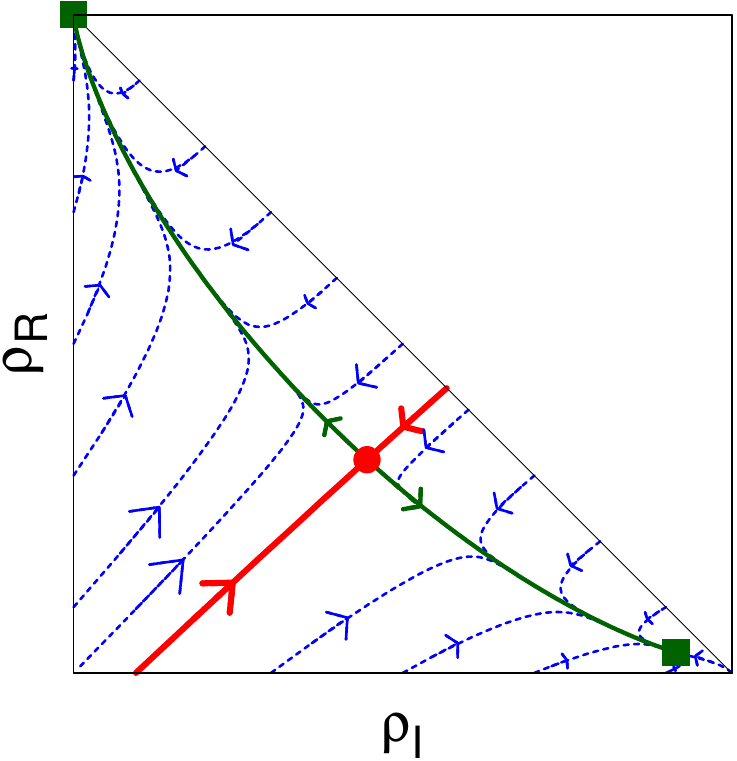} 
\caption{(color online) Phase-space ($\rho_I$, $\rho_R$) for the NG
  model with biased population replacement illustrating fixed points,
  separatrix and phase flows (represented by blue dashed lines) 
  for (left) $n=0$ and (right)
  $n=0.025$. The green squares are the attractive stable solutions
  $(1)$ and $(2)$, while the red circles are the solution $(3)$ that
  correspond to a saddle point with an attractive and a repulsive
  direction. Only the fraction of the physical phase space with
  $\rho_I+\rho_R < 1$ is represented.}
\label{phase_flow} 
\end{figure}

The physical stationary solutions are represented in
Fig.~\ref{stationary_solutions} as a function of $n$. The stability
of the generic solution $(\rho_I^*,\rho_R^*)$ as a function of $n$ is
investigated by looking at the eigenvalues and eigenvectors of the
stability matrix, defined through the equations: 

\be \frac{d}{dt}
\left[
\begin{array}{cc}
\delta \rho_I \\ \delta\rho_R
\end{array} 
\right]
=
\left[
\begin{array}{cc }
-r-\nu & 2\nu(\rho_R^* -1) \\ 
2\nu(\rho_I^* -1) & -r-\nu
\end{array}
\right]
\left[
\begin{array}{cc}
\delta \rho_I \\ 
\delta \rho_R
\end{array} 
\right].
\ee
The eigenvalues are given by: \be \lambda_{1,2}= \pm 2 \nu
\sqrt{(\rho_I^* -1)(\rho_R^*-1)} - (r+\nu). \ee
Figure~\ref{phase_flow} reports the complete phase flow in the space
($\rho_I$, $\rho_R$) for $n=0$ and $n=0.025$. For
$(\rho_I^{(1)},\rho_R^{(1)})=(0,1)$ both eigenvalues are negative: the
fully regular state is always attractive and stable and for $n >
(2\sqrt[3]{4}/3-1) \approx 0.0583$ it is the only physical solution.
For $n< (2\sqrt[3]{4}/3-1) \approx 0.0583$ the two other
physical solutions appear. ($\rho_I^{(2)}$, $\rho_R^{(2)}$) is always
stable and attractive and, for $n>0$, it always corresponds to states
for which $\rho_I+\rho_R < 1$. Let us now focus on the
($\rho_I^{(3)}$, $\rho_R^{(3)}$) solution. This solution corresponds
to a saddle point (the red circles in Fig.~\ref{phase_flow}) since it
has one positive and one negative eigenvalue. The separatrix in the
attractive direction corresponds to the eigenvector associated to the
negative eigenvalue: 

\be \rho_R = \rho_R^{(3)}+ m(\rho_I - \rho_I^{(3)}) \ee
\noindent with $m=\sqrt{ (1-\rho_I^{(3)}) /\sqrt{(1+n)\rho_I^{(3)}}}$
(the thick red solid line in the figure). The other separatrix is locally
approximated in the neighborhood of ($\rho_I^{(3)}$, $\rho_R^{(3)}$),
by 
\be \rho_R = \rho_R^{(3)}- m (\rho_I - \rho_I^{(3)}) \ee
\noindent (the thin green solid line in Fig.~\ref{phase_flow}). 
In Appendix A we report the explicit expressions for the limit case
$n=0$, i.e., the original Naming Game.
The model exhibits a discontinuous transition between a phase (high
values of $n=r/\nu$) where, whatever the initial condition, the
stationary state is fully regular and a phase (low values of $n$)
where both the fully regular state (solution $(1)$) and a state with a
large fraction of irregulars (solution $(2)$) are stable. As depicted
in Figure~\ref{phase_flow}, the initial condition determines which one
of the two states is asymptotically reached. In particular if the
initial condition is above the separatrix corresponding to the
attractive direction (thick red line) all individuals converge to the fully
regular state (solution $(1)$); on the other hand if the initial
condition is below that separatrix the system converges to the
solution $(2)$ where a fraction $\rho_I^{(2)}$ ($\rho_I^{(2)}=1$ for
the case with no replacement, $n=0$) of irregulars coexists
dynamically with regular and mixed individuals. Only initial
conditions exactly on the separatrix (thick red line) lead to convergence
to solution $(3)$. The predictions of the MF theory are confirmed by
numerical simulations of the actual agent-based model
(Fig.~\ref{stationary_solutions}).

In summary, the Naming Game with biased population replacement
exhibits a discontinuous transition as a function of $n$. The
discontinuity also implies a dependence of the final steady state on
the initial condition which provides a theoretical justification for
the observation of a range of frequencies where both fully regular and
mostly irregular verbs exist (see Fig.~\ref{figplos}).
For every frequency in this interval,
verbs will converge to the fully regular state $(1)$ or to the mostly
irregular state $(2)$, depending on the initial values of $\rho_R$ and
$\rho_I$. This phenomenology is in agreement the empirical findings
reported in~\cite{cuskley_2014} of the existence of a discontinuous
transition between regular and irregular forms as a function of the
frequency of usage.

\subsection{Model CT:  A three-state model with a continuous transition}
We now consider a model with the
interaction rules presented in Table~\ref{3stateRules}, columns CT,
which differs from the NG case essentially because the hearer never discards 
the mixed state.
The mean-field equations for the
evolution of the densities in this case are written as:
\begin{widetext}
\begin{equation}
\left\{
\begin{array}{lll}
\dot{\rho}_I &=& - r \rho_I + \nu \left\{- \rho_I \rho_R + \frac{1}{2}
\rho_R [1-(\rho_I+\rho_R)] + \frac{1}{2} [1-(\rho_I+\rho_R)]^2 \right\} \\
\dot{\rho}_R &=&  r (1-\rho_R) + \nu \left\{- \rho_I \rho_R + \frac{1}{2}
\rho_I [1-(\rho_I+\rho_R)] + \frac{1}{2} [1-(\rho_I+\rho_R)]^2 \right\}.
\end{array}
\right.
\end{equation}
\end{widetext}
By imposing the stationarity condition $\dot{\rho}_I =
\dot{\rho}_R = 0$ and summing and subtracting the two equations it is
possible to reduce them to
\begin{equation}
\left\{
\begin{array}{l}
r(1-y)-\frac{\nu}{2} y(1-x)=0\\
r(1-x)+\frac{\nu}{2} \left(y^2-3x+2\right)=0,
\end{array}
\right.
\end{equation}
where we have introduced the auxiliary variables 
$x=\rho_R+\rho_I$ and $y=\rho_R-\rho_I$.
From the second equation one obtains $x=(n+1+y^2/2)/(n+3/2)$.
\begin{figure} 
\includegraphics*[width=\columnwidth]{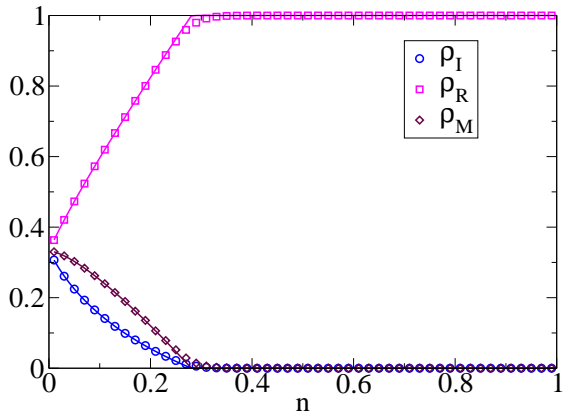}
\caption{(color online) Stationary value of the fraction of irregulars
  $\rho_I$, regulars $\rho_R$ and mixed $\rho_M$, as a function of $n$, 
  for the rules of model CT. Results of numerical simulations (symbols) 
  are compared with the analytical predictions (solid lines). The rounding
  of the transition in simulation results is due to finite size effects.}
\label{3stateContinuous} 
\end{figure}
Inserting this expression into the first equation, one is left with a
third order algebraic equation, which always admits three real solutions.
One of them is, for any $n$, the fully regular solution $(x,y)=(1,1)$.
The two others are always unphysical (being larger than 1 or smaller than -1) 
except for $n \le n_c=(\sqrt{17}-3)/4 \approx 0.2807$. 
In such a case another physical solution appears
\be
y=2\sqrt{-Q} \cos\left(\frac{\theta+4\pi}{3} \right),
\ee
where $Q=[2(n+1)-(2n+1)(2n+3)]/3$, $R=-n(2n+3)$ and $\theta=\cos^{-1}[R/\sqrt{-Q^3}]$.
The expressions for $\rho_R$ and $\rho_I$ are obtained from
the relations $\rho_R= (x+y)/2$ and $\rho_I= (x-y)/2$. 
The physical stationary solutions are reported in Fig.~\ref{3stateContinuous},
along with the results of the numerical simulations of the agent-based
model that well fit the theoretical predictions. 
The plot illustrates that the transition occurring at
at $n_c=(\sqrt{17}-3)/4$ is continuous.

\subsection{Model NT: A three-state model without a transition}

Let us now focus on another set of rules, reported in
Table~\ref{3stateRules}, column NT. Consider the case in which an
individual in the mixed state M is undecided about which one of R and
I is acceptable, and let the interactions between speaker and hearer
be symmetric. When an individual in a state I interacts with one in a
state R both become confused on which form is right and therefore both
switch to M. When an individual in a state M interacts with one in a
state I or R, the outcome of the interaction depends on which form the
individual in the mixed state uses: if it is the same one used by his
partner (with probability 1/2) then nothing changes, if it is the
alternative one (with probability 1/2) then the partner becomes
confused and also switches to the mixed state M.  Interactions among
individuals both in the same state I or R do not produce any
change. The outcome of an interaction among two individuals in a state
M depends on which form they use: if they both use the regular
inflection (with probability 1/4) they both switch to status R, if
they both use the irregular inflection (probability 1/4) they both
switch to status I, if they use different inflections (probability
1/2) they both remain in the mixed state M, coherently with the rules
regulating the outcomes of the previous interactions.

It is easily seen that under this rule set the system converges to the
fully regular state for any value of the ratio $r/\nu$. The MF
equations are:
%
%\begin{widetext}
\begin{equation}
\left\{
\begin{array}{lll}
\dot{\rho}_I &=& - r \rho_I + \nu \left\{-2 \rho_I \rho_R + \frac{1}{2}
 [1-(\rho_I+\rho_R)]^2  \right\} \\
\dot{\rho}_R &=&  r (1-\rho_R) + \nu \left\{-2 \rho_I \rho_R + \frac{1}{2}
[1-(\rho_I+\rho_R)]^2  \right\}.
\end{array}
\right.
\end{equation}
%\end{widetext}
%
Summing and subtracting the first equation from the second, 
 one obtains
\begin{equation}
\left\{
\begin{array}{l}
\dot{y}/\nu=n(1-y)\\
\dot{x}/\nu=n(1-x)-2x+y^2+1
\end{array}
\right.
\label{NT}
\end{equation}
 which shows that, for $n>0$ the fully regular state $\rho_R=1$, $\rho_I=0$
is the only possible stationary state. The conclusion is that, for any
value of $\nu$ (and $n>0$), this set of rules always leads to a
completely regular state for all individuals. 

Unlike the previous models this one is discontinuous in the limit
$n\rightarrow 0$: the model with replacement does not converge in 
the limit of vanishing replacement to the model with $n=0$. It is easy
to see that in the case $n=0$ from Eq.~(\ref{NT}) $\dot{y}\equiv 0$;
therefore the unbalance $y_0$ between $\rho_I$ and $\rho_R$
present in the initial condition is preserved during the dynamics, while
$x$ converges to $(1+y_0)/2$. Therefore the stationary state is
continuously dependent on the initial condition, and given by
$\rho_R=(1+y_0)^2/4$, $\rho_I=(1-y_0)^2/4$, which gives the fully
regular solution only as long as the system is initiated in the fully
regular state.

To conclude this section,  we observe that
in three--state models different microscopic interaction rules give
rise to qualitatively different behaviors. In the next section we
present a general approach to the modeling schemes presented so far,
clarifying why they give rise to different phenomenologies.

\section{General theory}

In this section we present a very general three--state model that
provides a unified framework for generic sets of interaction rules,
and we solve it analytically within the mean-field approximation. This
framework allows us to comprehend the origin of the different
behaviors found in the specific models investigated in the previous
sections, providing a complete understanding of the global
phenomenology of three-state models. We start by considering a general
two--state model first, as this elucidates why the more complex
three--state model is needed and how it behaves. We then consider a
very general three-state model, encompassing all models considered
before as particular cases. This approach will clarify a number of
general points. In particular it will show how the nature of the 
transition for both two and three--state models depends on the microscopic
rules, and clarify the role of asymmetries in the
behavior of the speaker and hearer in the communication process.

\subsection{General two--state model}
\label{General2}
Each individual is either in state R or I. At rate $r$ each individual
is replaced by one in state R. At rate $\nu$ an interaction occurs
among two randomly selected individuals, the speaker and the hearer.
We indicate the state of the pair of individuals in interaction as
(X,Y), where X is the state of the speaker and Y of the hearer. As
reasonable, we assume that nothing happens if the two individuals are
in the same state [(R,R) $\rightarrow$ (R,R), (I,I) $\rightarrow$
(I,I)]. We first consider the case of deterministic rules, i.e., the
state at the end of the interaction is fully determined by the initial
state. Starting with the state (R,I) we parametrize the interaction
rule by means of the coefficient $\gamma_{RI}$, which gives the
variation in the number of individuals in state I. For example, for
the interaction [(R,I) $\rightarrow$ (R,R)] $\gamma_{RI}=-1$, while
for [(R,I) $\rightarrow$ (I,I)] $\gamma_{RI}=1$. Analogously, when the
initial state is (I,R) the rule is parametrized by $\gamma_{IR}$.

The mean field equation for this process is simply
\be 
\dot{\rho}_I = - r \rho_I + \nu (\gamma_{RI}+\gamma_{IR})\rho_I (1-\rho_I).
\label{2stateGT}
\ee Obviously $\rho_R=1-\rho_I$. It follows immediately from
Eq.~(\ref{2stateGT}) that assuming distinct asymmetric roles between
speaker and hearer has no effect whatsoever on the collective
behavior, since only the cumulative coefficient
$\gamma=\gamma_{RI}+\gamma_{IR}$ enters the equation. The distinction
between hearer and speaker is therefore {\em irrelevant} and any model
defined by an asymmetric set of rules behaves exactly as its
symmetrized version. This observation allows to specify any two-state
model by means of just one parameter, $\gamma$, with values between -2
and 2.

The general solution of Eq.~(\ref{2stateGT})
is obtained by replacing $n$ with $2n/\gamma$ in Eqs.~(\ref{caseA})
and~(\ref{caseAstationary}).  The sign of $\gamma$ determines the
nature of the transition: for
$\gamma>0$ (rules biased in favor of I) there exists a continuous
transition with $n_c=\gamma$, while for $\gamma\leq 0$ (rules unbiased
or biased against I) there is no transition, and the fully regular
state is the only stationary solution.  
Therefore we conclude in general that the system is driven towards 
a fully regular state unless a bias in the interactions 
compensates for the increase in the R population due to replacement.

In the most general case, the outcome of each interaction is
decided probabilistically.
In this case $\gamma_{IR}$ and $\gamma_{RI}$ are defined as the {\em average}
increase in I states in the interaction, each of them assuming any real value 
in $\left[-1,1\right]$. Correspondingly $\gamma$ can assume any real
value in $[-2,2]$. It is immediate to realize that the dynamics is again 
described by Eq.~(\ref{2stateGT}) and all the above
conclusions hold.

\subsection{General three--state model}

As demonstrated explicitly in the case of two--state models,
asymmetric interaction rules (such as those in
Table~\ref{3stateRules}) produce exactly the same behavior as their
symmetrized version also in three--state models.  It is therefore possible
to express all possible interaction rules among individuals in a way
analogous of what we have done for the two--state case. Let us define
$\gamma_i$, $\phi_i$, and $\delta_i$ as the average variation in the
number of individuals in state I, R or M respectively, occurring when
an interaction of type $i$ takes place.  We will denote with $i=1$ the
interactions with initial state (I,R) and (R,I), $i=2$ for (R,M) and
(M,R) , $i=3$ for (I,M) and (M,I) , $i=4$ for (M,M).  As above, we are
assuming that no change occurs when two individuals in state R (or two
individual in state I) interact.  The interaction is instead
nontrivial in the case (M,M).  Conservation of the number of
individuals implies $\gamma_i+\phi_i+\delta_i=0$, for any $i=1,2,3,4$,
which reduces the number of independent parameters from 12 to 8.
Notice that these quantities may be non--integer when we allow for
different possible final states from a given initial state (each with
a given probability).

With this parametrization of the dynamical rules we can write the
rate equations for the evolution of the system in the most general
case:
\begin{widetext}
\be
\left\{
\begin{array}{lll}
\dot{\rho}_I/\nu &=& - n \rho_I +  \left\{2 \gamma_1 \rho_I \rho_R + 2 \gamma_2
\rho_R [1-(\rho_I+\rho_R)] + 2 \gamma_3 \rho_I [1-(\rho_I+\rho_R)]
+ \gamma_4 [1-(\rho_I+\rho_R)]^2 \right\} \\
\dot{\rho}_R /\nu &=&  n (1-\rho_R) +  \left\{2 \phi_1 \rho_I \rho_R + 2 \phi_2
\rho_R [1-(\rho_I+\rho_R)] + 2 \phi_3 \rho_I [1-(\rho_I+\rho_R)]
+ \phi_4 [1-(\rho_I+\rho_R)]^2 \right\}
\end{array}
\right.
\label{rateeq}
\ee
\end{widetext}
where $n=r/\nu$ is the rate of the replacement process relative to the
frequency of interaction.

We now focus on the case of unbiased interactions that do not favor
either the regular or the irregular form of the verb.  In other words
the interaction rules are perfectly symmetric under the exchange
between R and I; the only mechanism that breaks the symmetry between
the regular and the irregular form is replacement, which favors
the diffusion of the former. 
This choice is based on the observation that, while in the two-state 
case unbiased interaction rules always drive the system towards
the fully regular state, the existence of a mixed
state allows the survival of the irregular form even for some R--I
symmetric interactions.  This can be deduced from the three models
presented in the previous section, which all have unbiased rules
yet exhibit three qualitatively different behaviors.
Hence symmetric interaction rules are general enough to lead to
continuous or discontinuous transitions or to the absence
of any transition.

The assumption of unbiased interaction implies the following
additional relations among the parameters: \be \phi_1 =
\gamma_1~,~~~~\phi_2 = \gamma_3~,~~~~\phi_3 = \gamma_2~,~~~~ \phi_4 =
\gamma_4 \,,\label{unbiased} \ee thus reducing the number of free
parameters to 4 (we choose to use the four $\gamma_i$).
The values of these parameters are not arbitrary. 
An (R,I) interaction cannot produce an increase in the number of 
individuals in the I state, since the same change must occur also 
for individuals in the R state, since $\gamma_1=\phi_1$: hence 
$\gamma_1 \leq 0$. In the same interaction the number of irregulars 
cannot decrease by more than 1: $-1 \leq \gamma_1$. With similar
considerations it is not difficult to verify that
the $\gamma_i$ parameters are bounded as follows:
\be 
\!\!\!-1 \leq\! \gamma_1 \!\leq 0, \,\,\, 0 \leq \!\gamma_2 \!\leq
2, \,\,\, -1\leq \!\gamma_3 \!\leq 1, \,\,\, 0 \leq \!\gamma_4 \!\leq 1.  
\ee

Introducing the relations~(\ref{unbiased}) in Eqs.~(\ref{rateeq}) 
 it is possible to write Eqs.~(\ref{rateeq}) in a 
particularly simple form by defining the auxiliary quantities
$x=\rho_R+\rho_I$ and $y=\rho_R-\rho_I$, which represent the fraction
of individuals in an unmixed state, and the excess fraction of R states
with respect to I states, respectively: 
\be
%\begin{eqnarray}
\label{generic_both1}
\left\{
\begin{array}{l}
\dot{x}/\nu=a x^2 +c y^2 + 2 d x + f \\ 
\dot{y}/\nu=(2(\gamma_3-\gamma_2)-n)y-2(\gamma_3-\gamma_2)xy+n
\end{array}
\right.
%\end{eqnarray}
\ee
where 
\be
\begin{array}{llllllll}
a&=&\gamma_1 -2 (\gamma_2+\gamma_3) + 2 \gamma_4 & \,\,& &c& = &-\gamma_1 \\
d&=&(\gamma_2+\gamma_3)-2 \gamma_4-n/2  & & \,\,& f&=&2\gamma_4+n.
\end{array}
\ee 

Physically sensible solutions must be in the range 
$0\leq x \leq 1$ and $-x \leq y \leq x$.  
Notice that $c\geq 0$, $f \geq 0$, while the sign of $a$ and $d$ may vary
and the coefficients are related by 
\be
\label{constraint}
a+c+2d+f=0 \,.  
\ee 
Equation~(\ref{constraint}) implies that, for any
choice of the parameters, $(x,y)=(1,1)$ [which corresponds to the fully
regular state ($\rho_R=1,\rho_I=0$)] is always a stationary
solution of Eq.~(\ref{generic_both1}).
By imposing stationarity in Eqs.~(\ref{generic_both1}), other stationary
solutions 
can be determined.  For specific values of the parameters $\gamma_i$ it is
always straightforward to solve analytically for the stationary solutions
of Eqs.~(\ref{generic_both1})
(which boils down to the solution of a third-order algebraic equation)
and study their behavior as a function of $n$.  

In the following we derive instead in full generality conditions on
the parameters $\gamma_i$ for the existence, as a function of $n$, of
a continuous transition, a discontinuous one or no transition at all.

\subsubsection{Case $\gamma_3=\gamma_2$: No-transition}
Let us first consider the special class of models with
$\gamma_3=\gamma_2$.  In this case the second of Eqs.~(\ref{generic_both1})
trivially yields at stationarity $y=1$, giving the two solutions 
$(x_1,y_1)=(1,1)$ and $(x_2,y_2)=(-(2d+a)/a,1)$. 
The second solution is always unphysical.
Indeed, from Eq.~(\ref{constraint}) it follows that $-(2d+a)>0$.
Hence $x_2$ is smaller than 0 if $a<0$.
On the other hand, for $\gamma_3=\gamma_2$ the value of 
$\gamma_3+\gamma_2$ is always positive, since $\gamma_2 \ge 0$. 
This implies $-(2d+a)>a$ so that $x_2>1$ if $a>0$.
We conclude that when $\gamma_3=\gamma_2$ the only possible
stationary state is the fully regular one: no transition may occur. 
As shown in Table~\ref{Examples} the NT model considered in the previous
section falls in this class of models, featuring $\gamma_2=\gamma_3=0$.
We will see below that also the degenerate case $\gamma_1=0$ 
implies no transition, irrespective of the value of the other parameters. 
\begin{table}
%\begin{tabular}{c || c| c |c}
\begin{tabular}{l c c c}
\hline
\hline
&Model NT$\,\,\,$&  Model CT $\,\,\,$&	Model NG$\,\,\,$\\
\hline
$\,\,\,\,\gamma_1\,\,\,\,\,\,\,\,$& -1& -1/2, &-1/2\\
%\hline
$\,\,\,\,\gamma_2\,\,\,\,\,\,\,\,$&0&1/4& 0\\
%\hline
$\,\,\,\,\gamma_3\,\,\,\,\,\,\,\,$&0&0&1/2\\
%\hline
$\,\,\,\,\gamma_4\,\,\,\,\,\,\,\,$&1/2&1/2&1\\
\hline
\hline
\end{tabular}
\caption{The table summarizes the values of the set of $\gamma_i$ parameters
for the three models studied in the previous section. }
\label{Examples}
\end{table}

\subsubsection{Case $\gamma_3\neq \gamma_2$: Existence of a transition}
\begin{figure}
\includegraphics*[width=\columnwidth]{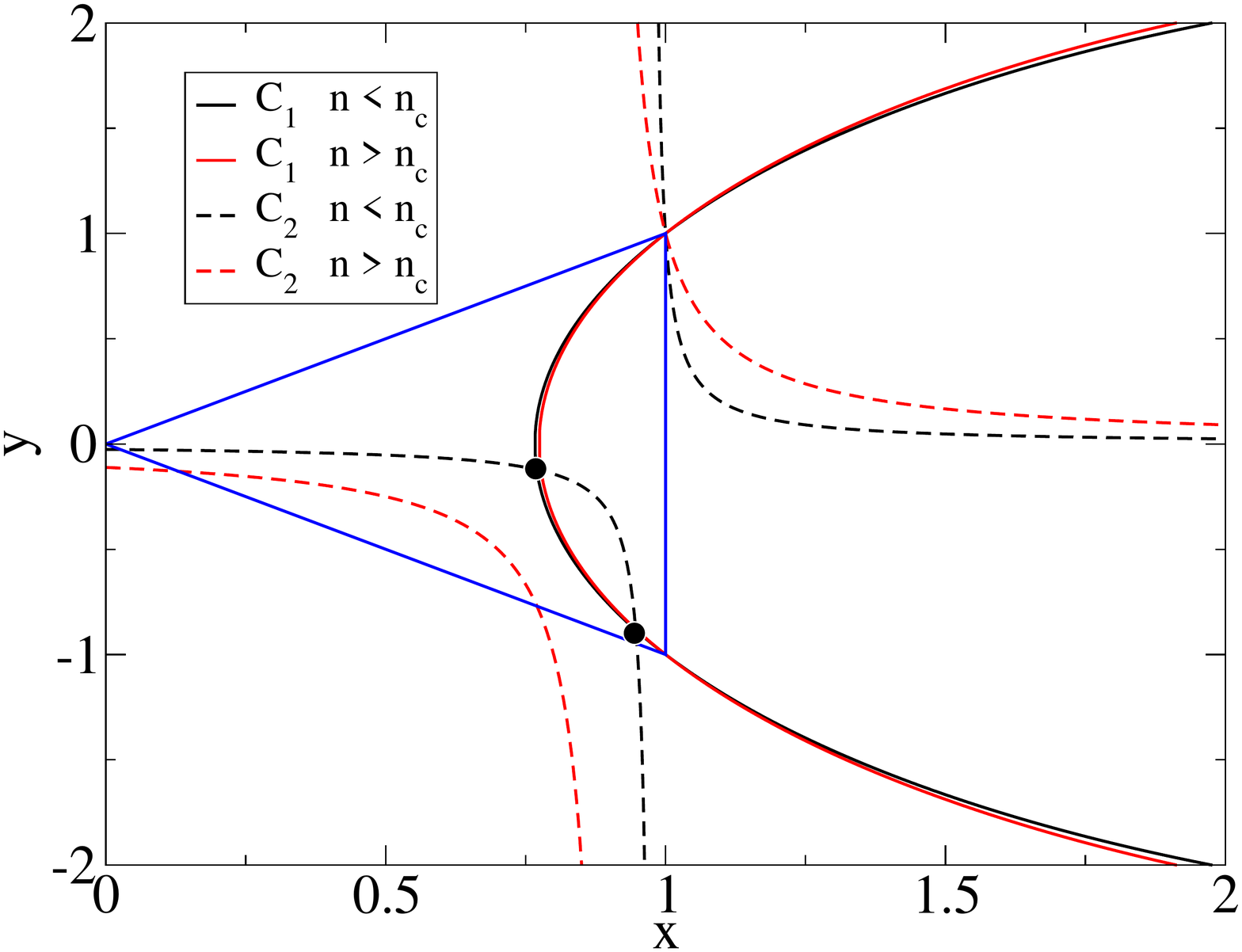}
\includegraphics*[width=\columnwidth]{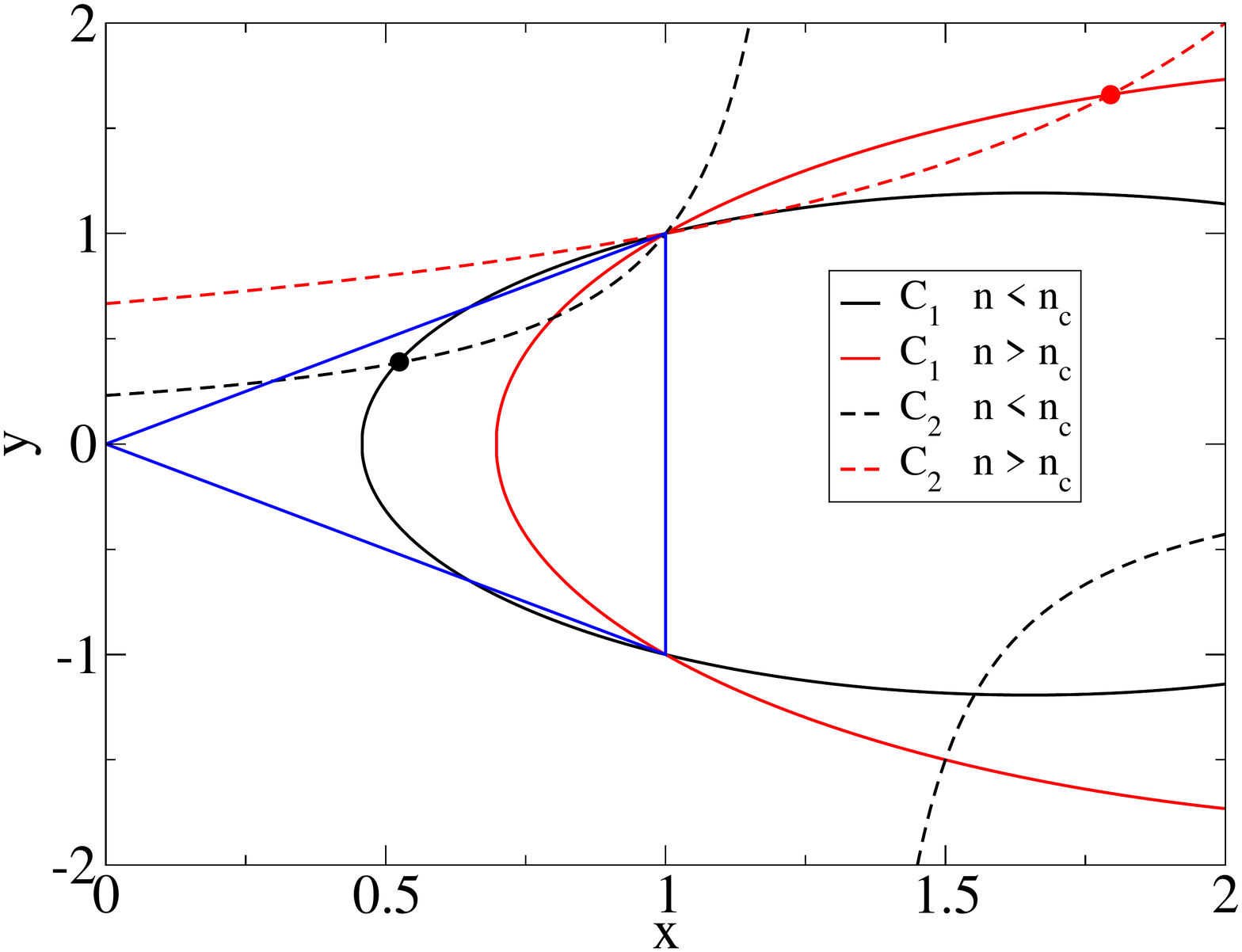}
\caption{(color online) 
  Plot of the two conic sections ${\cal C}_1$ and ${\cal C}_2$ for
  $\gamma_3>\gamma_2$ (top) and for $\gamma_3<\gamma_2$ (bottom),
  illustrating the intersections above [red (gray) curves] and below 
  (black curves)
  the critical $n_c$. The triangle is the region of physical interest.
  The circles denote the intersections other than the fully regular solution.}
\label{FigGeneric}
\end{figure}

Assuming now $\gamma_3 \neq \gamma_2$, and imposing stationarity in 
Eq.~(\ref{generic_both1}), we get 
\be
\label{generic}
\left\{
\begin{array}{lr}
a x^2 + c y^2 + 2d x + f = 0 & \,\,\,\,\,\,\left[{\cal C}_1\right]\\
y\left(x-(1-\epsilon)\right)=\epsilon  & \,\,\,\,\,\,\left[{\cal C}_2\right]
\end{array}
\right.
\ee
where  $\epsilon=n/[2(\gamma_3-\gamma_2)]$. 
The solutions of Eq.~(\ref{generic}) are given by the intersections between 
two conic sections ${\cal C}_1$ and ${\cal C}_2$. 
${\cal C}_2$ is an hyperbola with asymptotes $y=0$, and $x=1-\epsilon$. 
Depending on the sign of $\epsilon$, the two branches of the hyperbola 
lie in different quadrants with respect to the asymptotes
(see Fig.~\ref{FigGeneric}). 
In the limit  $\epsilon \rightarrow 0$, which corresponds to the case 
with no replacement, the hyperbola degenerates into the
pair of asymptotes, $y=0$ and $x=1$. 
The conic section ${\cal C}_1$ is an ellipse for $a>0$ and an hyperbola 
for $a<0$, turning into a parabola for $a=0$,
with axes in all cases parallel to the Cartesian coordinate system.
The intersections of ${\cal C}_1$ with the $y=0$ axis are, for $a \neq 0$,
\be 
x_1 = \frac{-d-\Delta}{a} \,,~~~~~ x_2 = \frac{-d+\Delta}{a}\,, 
\label{x1x2}
\ee where $\Delta=\sqrt{d^2-af}=\sqrt{(d+f)^2+cf}>0$. It is not
difficult to show that $x_1$ is always physical (between 0 and 1)
while $x_2$ is always unphysical (see Appendix B). In particular,
$x_2>1$ if $a>0$ and $x_2<0$ if $a<0$. In the limit $a\rightarrow 0$,
$x_1 \rightarrow -f/(2d)$ (which is the intersection point for the
parabolic case $a=0$) while $x_2 \rightarrow \pm \infty$, depending on
the sign of $a$. Hence, despite the change in the global behavior for
different values of $a$, ${\cal C}_1$ has always a similar shape in
the region of physical interest $0 \le x \le 1$: it crosses the $y=0$
axis for $x=x_1$ and is concave towards the right, passing through the
two points $(x,y)=(1,1)$ and $(x,y)=(1,-1)$ for any value of the
parameters. Notice that also ${\cal C}_2$ always goes through the
point $(x,y)=(1,1)$, so that
the fully regular state is always a stationary solution.

We now investigate in full generality the possible existence of other
stationary solutions, i.e. other intersections in the physical region.
Only the case $c=0$ ($\gamma_1=0$) needs to be treated separately, because
the conic section ${\cal C}_1$ degenerates into a pair of lines:
$(x-1)(x+(2d+a)/a)=0$, one ($x=1$) on the boundary of the physical
region, the other outside it. 
The fully regular solution $(x,y)=(1,1)$ is then the only stationary state,
and there is no transition.
We will assume $c \neq 0$ in what follows.

\paragraph*{$\gamma_3>\gamma_2$: Discontinuous transition.}
For $\gamma_3>\gamma_2$ ($\epsilon>0$), the hyperbola ${\cal C}_2$
lies in the upper--right and lower--left quadrants with respect to
the asymptotes (see Fig.~\ref{FigGeneric}).
Notice that the vertical asymptote is always at $x<1$.
Apart from the fully regular solution, one intersection occurs always
for $x<0$ or $x>1$ and is thus unphysical. Two other intersections may instead
occur between  ${\cal C}_1$ and the lower--left branch of ${\cal C}_2$.
For large $n$ these intersections do not exist as it can be recognized by
observing that for $\epsilon>1$ the lower branch of  ${\cal C}_2$ has
$x<0$ while ${\cal C}_1$ has $x>x_1>0$. 
However, when $n \to 0$, ${\cal C}_2$ shrinks towards its 
asymptotes $x=1-\epsilon$  and $y=0$, while ${\cal C}_1$ does not change much. 
At some critical value $n=n_c$, ${\cal C}_2$ starts intersecting 
${\cal C}_1$, so that for $n<n_c$ there are two intersections 
(see Fig.~\ref{FigGeneric}), both in the physical region because
the derivative of ${\cal C}_1$ for $x=1$ is $-[1+(d+f)/c]<-1$.
In this case the system undergoes a discontinuous transition at $n_c$. 
Notice when these two solutions exist they are in the region 
$y \le 0$, implying that $\rho_I \ge \rho_R$, i.e., 
the fraction of individuals  in the I state is larger 
than the fraction of those in the R state.
As indicated in Table~\ref{Examples}, the model NG, which
features $\gamma_3-\gamma_2=1/2$, belong to this class and 
this explains why it undergoes a discontinuous transition.

\paragraph*{$\gamma_3<\gamma_2$: Continuous transition.}
For $\gamma_3<\gamma_2$ ($\epsilon<0$), the hyperbola ${\cal C}_2$
lies in the upper--left and lower--right quadrants
(see Fig.~\ref{FigGeneric}).  The lower branch
has $x>1-\epsilon>1$ and hence is unphysical. 
Therefore at most two of the four solutions (the intersections of
the upper branch with ${\cal C}_1$) are physical.
One of them is the fully regular state. 
To investigate the location of the other intersection one can
compare the slope of the two conic sections for $x=1$.
If the slope of ${\cal C}_1$ is larger than the slope of ${\cal C}_2$
(which happens for large $n$)
the second intersection occurs for $x>1$, it is unphysical and as
a consequence the fully regular state is the only stationary solution.
If $n$ is reduced the slope of ${\cal C}_1$ at $x=1$ decreases
while the slope of ${\cal C}_2$ grows. 
At a critical value $n_c$ the two slopes are equal and for $n<n_c$ 
the second intersection becomes physical ($x<1$).
We conclude therefore that the system undergoes a continuous
transition between a fully regular state and a state with coexisting
regular and irregular individuals. 
Notice that in this case, since the physical intersections
have $y \ge 0$, necessarily $\rho_I \le \rho_R$.  
The value of $n_c$ is easily determined by the condition that the two
slopes are equal, and turns out to be
\be 
n_c =
(\gamma_1\!-\!\gamma_2\!-\!\gamma_3)\! \pm\!
\sqrt{(\gamma_1\!-\!\gamma_2\!-\!\gamma_3)^2+4 \gamma_1
  (\gamma_3-\gamma_2)}
\label{nc}
\ee 
which has always one positive value (the other being
always negative), coherently with the fact that there is always a
transition.  Remarkably, the value of $n_c$ does not depend at all 
on the coefficient $\gamma_4$, regulating the M-M interaction.
The model CT in the previous section has $\gamma_3=0$, $\gamma_2=1/4$ and
$\gamma_1=-1/2$ (see Table~\ref{Examples}). 
These values explain why it undergoes a continuous transition at $n_c=(\sqrt{17}-3)/4$.

\subsubsection{Stability analysis}
So far we have shown that below some critical value of $n$ additional
stationary solutions appear, beyond the fully regular solution.
To complete the demonstration of the existence of phase-transitions
we must analyze their stability.

By linearizing Eqns.~(\ref{generic_both1}) around the solution $(x^*,y^*)$
one gets:
\be \label{StabilityAnalysis}
\!\frac{1}{2\nu}\frac{d}{dt} \!\left[
\begin{array}{cc}
\delta x \\ 
\delta y
\end{array} \right]\!
=\!
\left[ \begin{array}{cc }
a x^{*}+d  &  c y^{*}\\ 
-(\gamma_3-\gamma_2) y^{*} & (\gamma_3-\gamma_2) (1-x^{*})-\frac{n}{2}
\end{array} \right]\!
\left[ \begin{array}{cc}
\delta x  \\ \delta y
\end{array}  \right].
\ee
For the fully regular state $(x^*,y^*)=(1,1)$
the eigenvalues of the stability matrix $M$ can be evaluated 
explicitly yielding
\be
\lambda_{1,2}=[\operatorname{tr}(M)\pm\sqrt{\Delta_1}]/2
\ee
where the trace of the matrix is
$\operatorname{tr}(M)=\gamma_1-\gamma_2-\gamma_3-n$ and
$\Delta_1=(\gamma_1+\gamma_2+\gamma_3)^2-8\gamma_1\gamma_2\geq 0$. 
Notice that $\Delta_1$ is independent of $n$ and positive,
implying that the two eigenvalues are always real. 
\begin{figure*} 
\includegraphics*[width=\columnwidth]{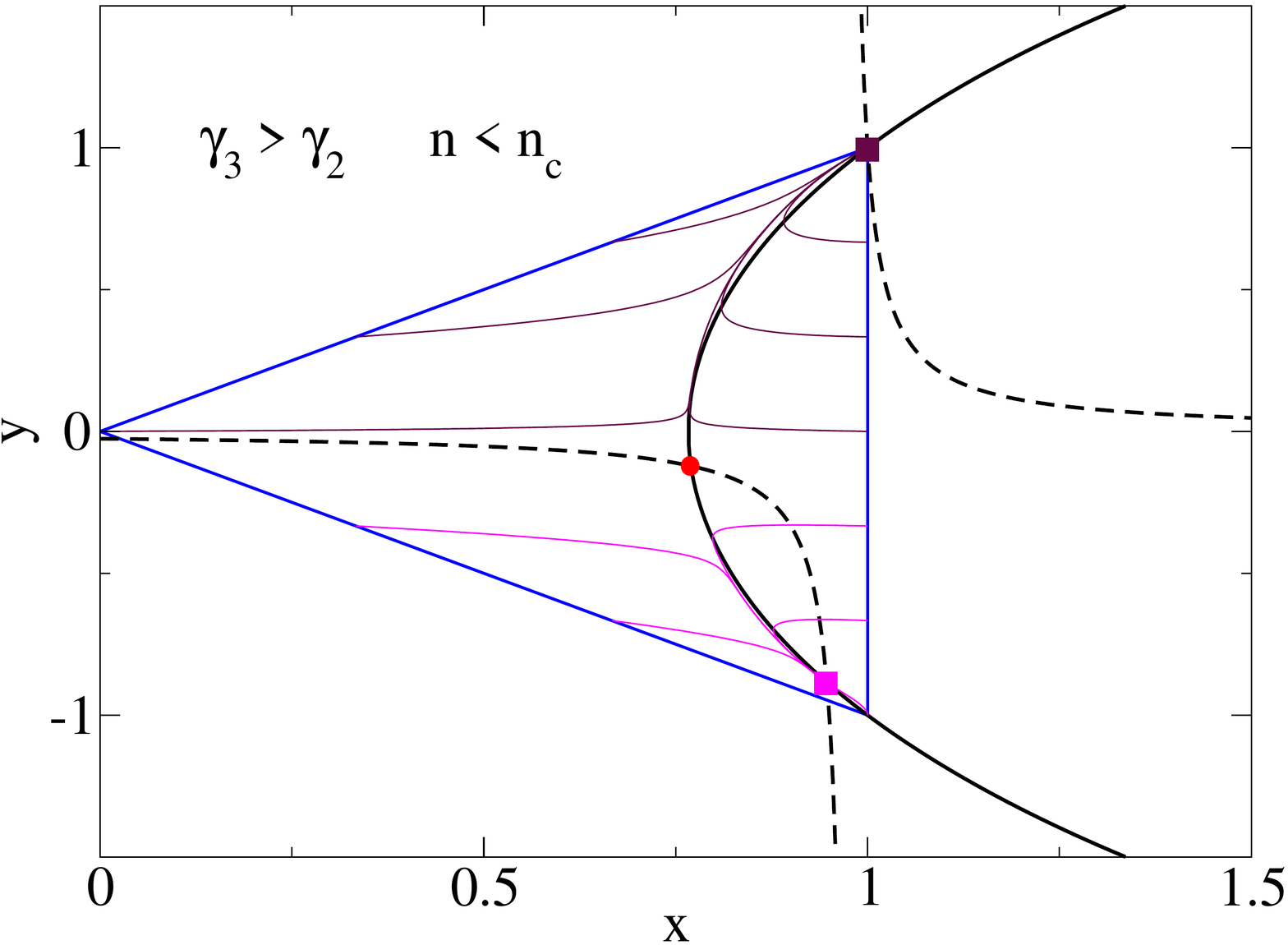} 
\includegraphics*[width=\columnwidth]{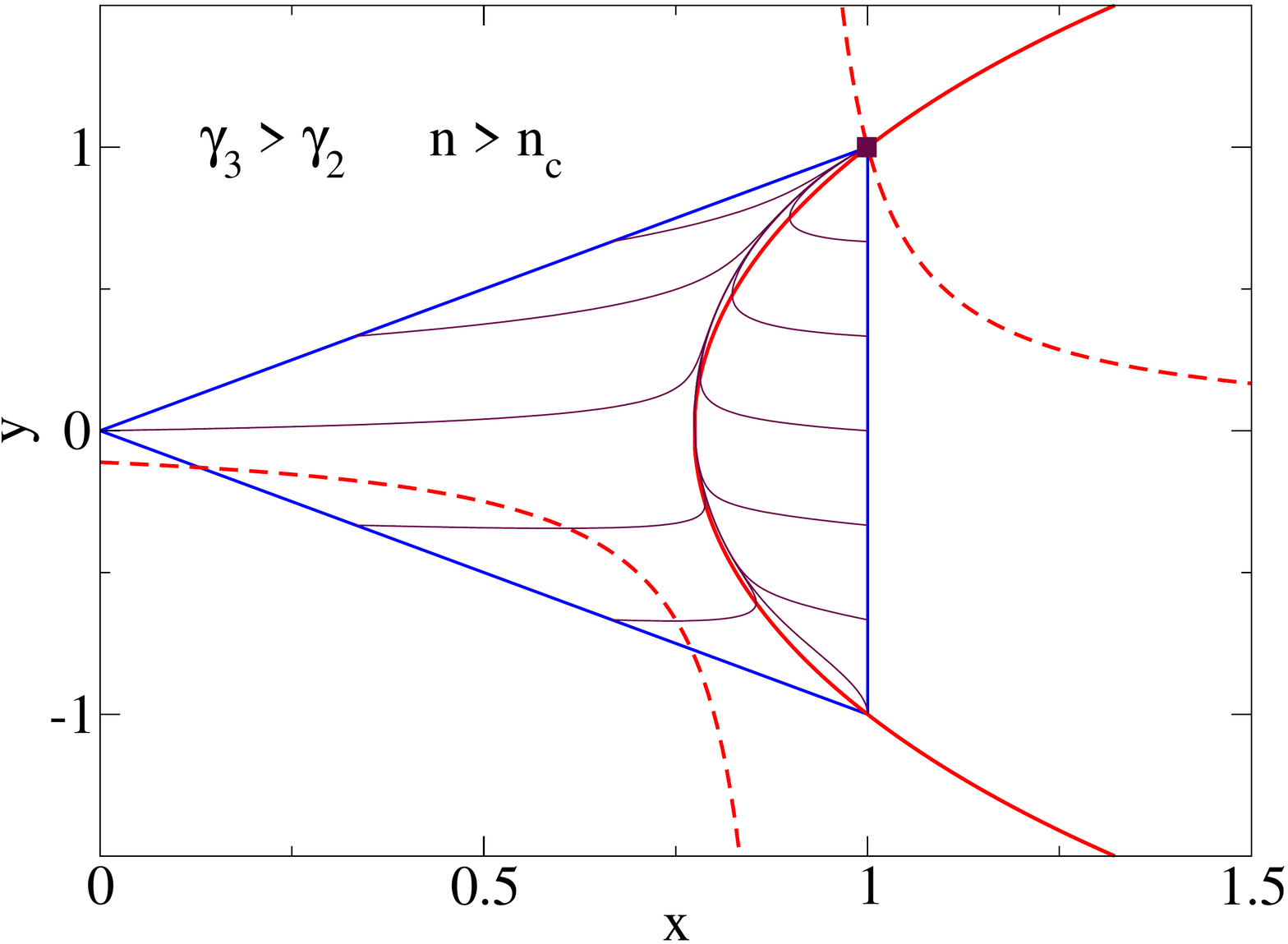} 
\includegraphics*[width=\columnwidth]{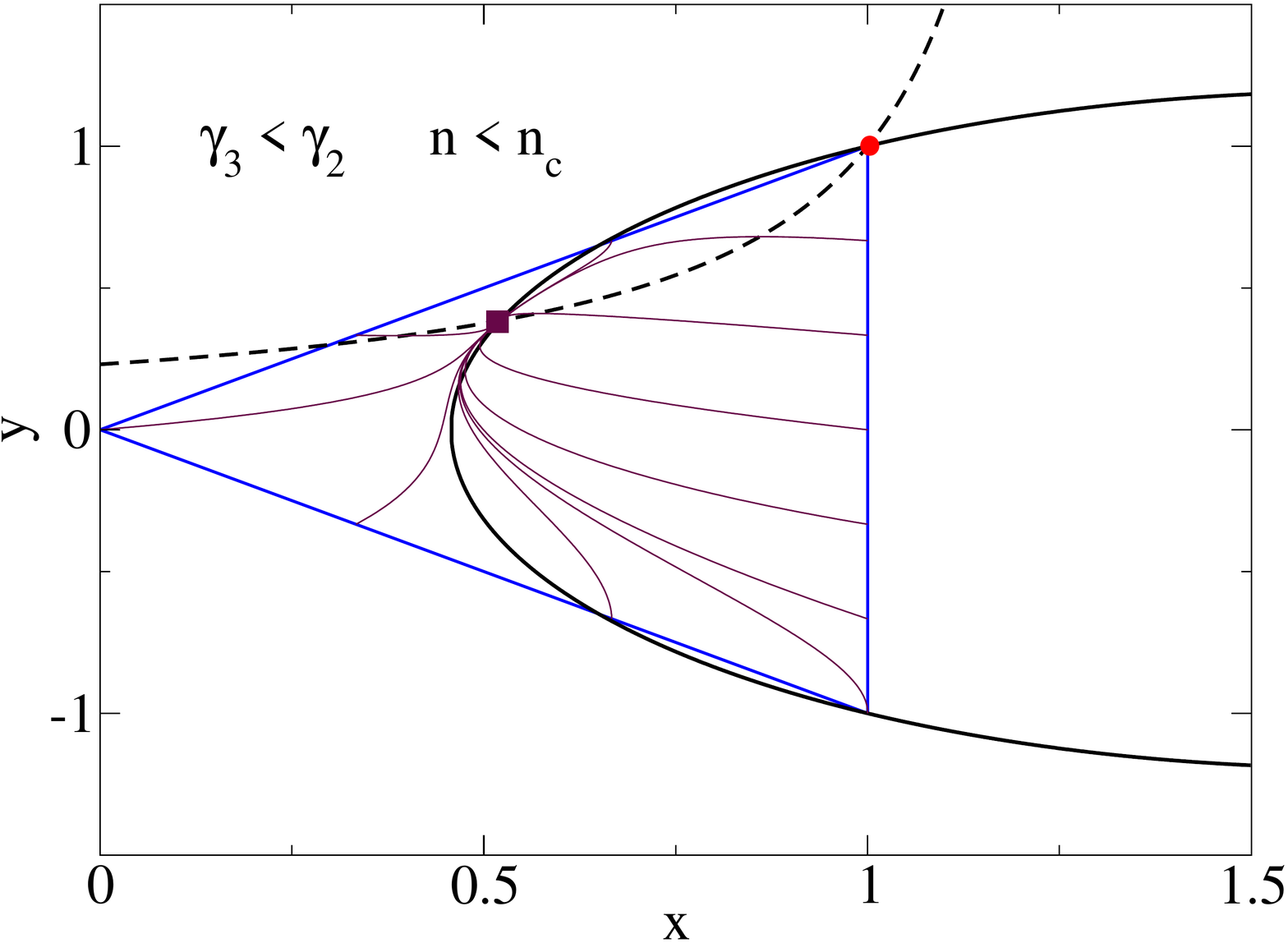} 
\includegraphics*[width=\columnwidth]{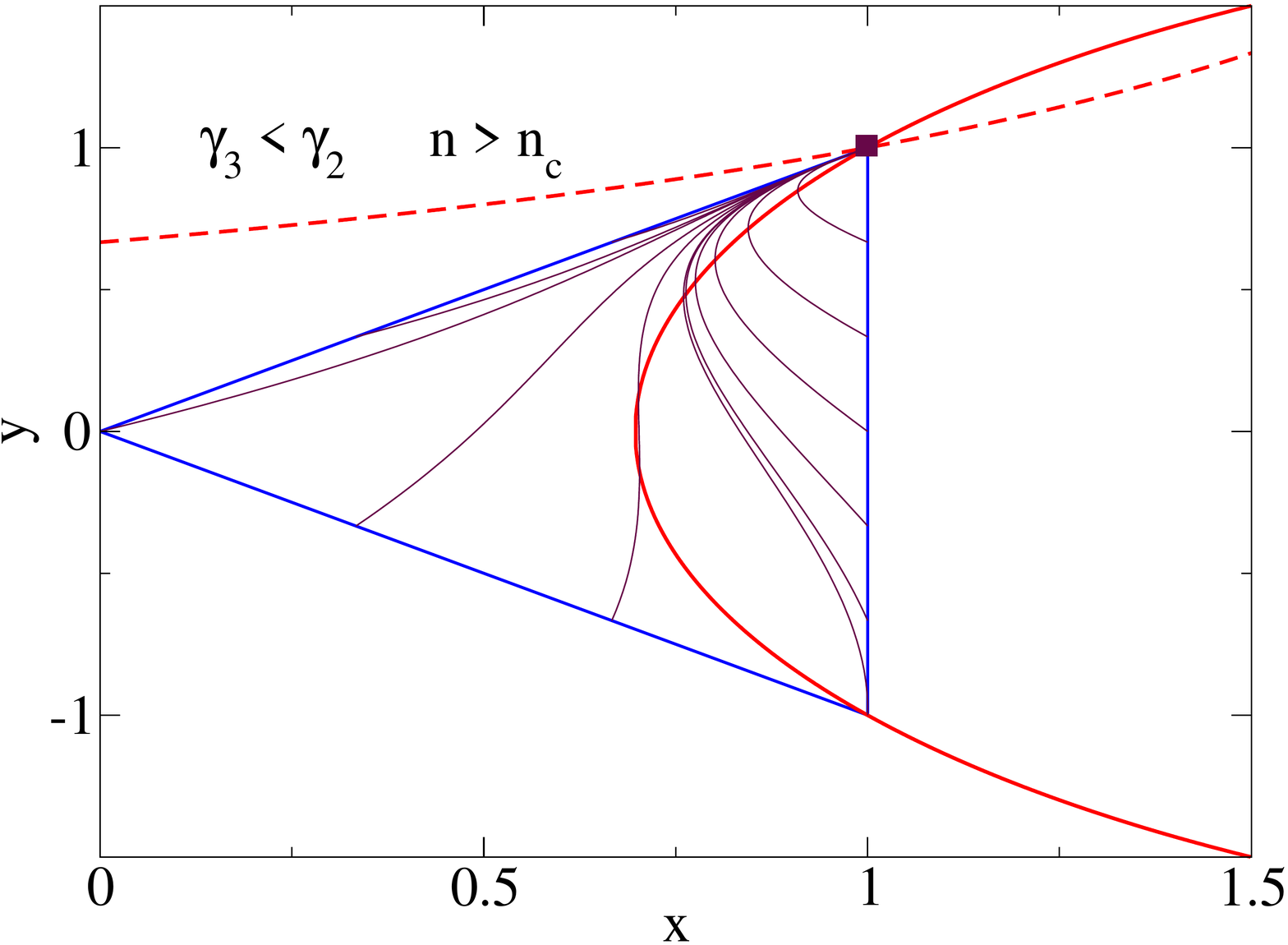} 
  \caption{(color online) 
    Phase-flow in the $(x,y)$ space for discontinuous (top) and
    continuous (bottom) transition both below (left) and above (right)
    the transition.  The triangle is the region of physical
    interest.  The thick curves are ${\cal C}_1$ (solid) and ${\cal
      C}_2$ (dashed).  Symbols indicate stationary solutions. The red
    circles are saddle points with an attractive and a repulsive
    direction. The squares are attractive stable solutions.  Thin
    solid lines are flow lines converging to the stable solution
    denoted by symbols with the same color (shade).}
\label{phase_space} 
\end{figure*}

In the case $\gamma_3>\gamma_2$, both eigenvalues are negative, as can
be deduced by considering that $\operatorname{tr}(M)$ (the sum of the
eigenvalues) is always negative (being the sum of four negative terms)
and $\det(M)=n^2/4-n(\gamma_1-\gamma_2-\gamma_3)/2-\gamma_1
(\gamma_3-\gamma_2)$ (the product of the eigenvalues) is always
positive for $n>0$. This last condition can be understood by
considering that $\det(M)$ is a parabola with upward concavity and
zeros for negative values of $n$ [given by Eq.~(\ref{nc})]. Hence, the
fully regular state is always stable. For small $n$, the two other
solutions appearing in the physical region are a saddle and a stable
fixed point [see Fig.~(\ref{phase_space})]: a discontinuous transition
occurs.

In the case $\gamma_3<\gamma_2$, corresponding to a continuous
transition, $\det(M)$ is positive for large $n$, but it changes sign
for $n<n_c$ [where $n_c$ is the only positive determination of 
 Eq.~(\ref{nc})] thus
implying that one of the eigenvalues becomes positive. Hence, below
the transition the fully regular state corresponds to a saddle, while
the other solution appearing in the physical region is stable [see
Fig.~(\ref{phase_space})].

Finally, we remark that the parameters corresponding to the Naming
Game, considered in the previous section as an example of models with
discontinuous transition, give $\Delta_1=0$, corresponding to two
degenerate eigenvalues, and only one eigenvector. In this case the
fully regular state is stable but {\it defective}: in principle
arbitrarily small changes of the parameters may result in a standard
stable fixed point if $\Delta_1>0$, or in a stable spiral when
$\Delta_1<0$. This last case would change the physical picture,
because spiralling trajectories would cross the boundary of the
physical domain before reaching the fixed point, and therefore the
physical system would be driven toward the boundary of the physical
region without reaching the fixed point. However, this possibility is
ruled out by the condition $\Delta_1 \geq 0$ which holds for
physically sensible values of the parameters $\gamma_i$.

\subsubsection{Wrap-up of the theory for the general model}
The conclusions that follow from the theory presented in this section
are very general and simple. The existence and the nature of a
transition depends only on the sign of $\gamma_3-\gamma_2$. If
$\gamma_3>\gamma_2$, as the frequency increases a discontinuous
transition occurs between a fully regular state and a state with
coexistence of fully regular and mostly irregular inflections. If
$\gamma_3<\gamma_2$, the transition is instead continuous. If
$\gamma_3=\gamma_2$, no transition occurs and the system always
reaches a fully regular state.

It is important to notice that the condition $\gamma_3>\gamma_2$ has a
very natural interpretation in the context of language dynamics. It
simply means that (recalling that $\gamma_2=\phi_3$) an interaction
between an individual in state M and one in state I, will produce an
increment in the use of the irregular inflection larger than the
increment in the use of the regular inflection. The fact that this
asymmetry alone is sufficient to give rise to a discontinuous
transition is a strong indication of the relevance of these
theoretical modeling efforts for the interpretation of empirical data.

\section{Conclusions}

This work has presented an investigation of agent-based models aimed
at understanding the processes regulating the interplay between rules
and exceptions in language dynamics. In particular, the models aim to
investigate the observed behavior of verbs in natural
language. Corpus data from natural language points to the
existence of a discontinuous transition as a function of the frequency
of usage: high frequency items are highly irregular and low
frequency ones are regular, while in an intermediate frequency range
coexistence between the two behaviors is observed.

In the minimal models considered each agent is endowed with an
inventory, containing the possible inflections (regular or irregular)
of a lemma. Two processes have the potential to change agents'
inventories over time: interaction and replacement. In interaction,
individuals influence each other, adding or deleting forms from their
inventories according to a specific set of interaction rules. In
replacement events, agents are substituted by new ``child''
individuals, who are automatically biased towards the regular form by
being ``born'' with a regular inventory.

We analyze two classes of models. In the first one each individual may
store in the inventory only one of the two competing inflections,
either the regular or the irregular one.
Three-state models instead
integrate a mixed state, which represents an individual who finds both
the regular and irregular forms acceptable. We solve these models
analytically within the mean-field framework and confirm the results
by means of numerical simulations. We first focus on a few specific models,
including the Naming Game for language dynamics. The analysis reveals that
the the global phenomenology changes qualitatively depending on the
interaction rules: one can observe the absence of a transition with a
move to total regularity, a continuous transition, or a discontinuous
one. We then consider a very general three-state model,
encompassing all previous examples as special cases, which allows
the description of the previous models with a set of four minimal
parameters describing the interaction rules.
From this comprehensive approach several results follow: \\
({\em i)} Asymmetries in the influence of the speaker and of the
hearer in interaction do not play any
role in the collective behavior of the system.\\
({\em ii)} In two-state models the fully regular state is the only
attractor unless the interaction rules are biased in favour of the
irregular inflection; the three-state models have instead nontrivial
behavior even when the rules are unbiased.\\
({\em iii)} Allowing for a third state is crucial for the appearance
of a discontinuous transition that cannot arise in two--state
versions of the model.\\
({\em iv)} In three state models the quantity $\gamma_3-\gamma_2$ rules
the macroscopic behavior by changing the nature of the transition:
when $\gamma_3>\gamma_2$,
a discontinuous transition is observed to a state where irregular
inflection is prevalent ($\rho_I>\rho_R$); in the opposite case,
when $\gamma_3>\gamma_2$ a continuous transition is observed, to a state
with $0<\rho_R<\rho_I$; in the case $\gamma_3=\gamma_2$ there is no
transition and the fully regular state  is reached for any frequency.\\
({\em v)} In the case $\gamma_3>\gamma_2$, above the discontinuous transition
the steady state depends on the initial condition:
verbs with the same frequency can end up as fully regular or mostly irregular, 
similarly to what is observed in empirical data.\\
({\em vi)} In the context of language dynamics, the condition
$\gamma_3>\gamma_2$ is satisfied by sets of more plausible rules, so that a
discontinuous transition is to be expected.

This model provides a framework that could potentially be used to
consider additional, more complex aspects of rule dynamics in
language. In particular, empirical data shows that the growth of
language contributes to the expansion of
regularity~\cite{cuskley_2014}, since a core aspect of a linguistic
rule's utility is that it can be generatively applied to new forms
(e.g., the past tense of the neologism {\it selfie} is
uncontroversially {\it selfied}).
Our model considers a word's frequency to be static over time;
however, natural languages are living, and populations, vocabulary
sizes, and turnover rates are not static. Furthermore, there are other
cognitive mechanisms beyond child learner biases that may contribute
to regularity dynamics. General memory constraints may contribute to
the persistence of highly frequent irregular
forms~\cite{cuskley_evolang_2014}, and adult learners may possess
qualitatively different regularization biases from
children~\cite{hudson}. Moreover, the use of a disordered topology for
the pattern of interaction, as opposed to the homogeneous mixing
assumed by the MF approach, combined with the interaction among the
different lemmas in the agents' inventories may lead to different
patterns of regularization in frequency. Future models might also
consider another key aspect in the persistence of irregularity: the
notion that irregular forms are not always exceptions, but sometimes
constitute sub-rules \cite{yang} (e.g., \textit{foot-feet/goose-geese,
  sing-sang/ring-rang}). Our model provides a basic starting point
from which to consider the complex dynamics underlying temporal trends
of the rules that form the core of language.

Finally, it is very important to stress that while models and results
are presented in this paper in terms of linguistic rule dynamics, they
are fully general and apply to any system where individuals have three 
possible internal states and the population exhibits turnover. The results 
presented in this paper, and in particular the conditions determining 
the existence of a transition and its nature, may have strong implications 
not only for linguistic rules, but also for all those systems.

\section*{Acknowledgments}
This work was supported by the European Science Foundation as part of
the DRUST project, a EUROCORES EuroUnderstanding
programme: http://www.eurounderstanding.eu/
The funders had no role in study
design, data collection and analysis, decision to publish, or
preparation of the manuscript.

\section{Appendix}

\subsection{}

In this appendix we report the stability analysis for the Naming Game
without biased replacement, i.e., with $n=0$ as discussed in
Sec.~\ref{sec:NGBR}. For the case $n=0$ the stability matrix is
given by:
\be
\frac{d}{dt} \left[
\begin{array}{cc}
\delta\rho_I \\ \delta\rho_R
\end{array} 
\right]
=
\left[
\begin{array}{cc }
-1 & 2(\rho_R -1) \\ 
2(\rho_I -1) & -1
\end{array}
\right]
\left[
\begin{array}{cc}
\delta\rho_I \\ \delta\rho_R
\end{array} 
\right]
\label{eq:stabn0}
\ee

\noindent whose eigenvalues are:

\be
\lambda_{1,2}= \pm \sqrt{4 (\rho_I^* -1)(\rho_R^*-1)} -1,
\ee

\noindent where, as before, ($\rho_I^*$, $\rho_R^*$) indicates the
generic stationary solution. For ($\rho_I^{(1)}$, $\rho_R^{(1)}$) and
($\rho_I^{(2)}$, $\rho_R^{(2)}$) both eigenvalues are negative and the
solutions are both stable. For ($\rho_I^{(3)}\simeq 0.382$,
$\rho_R^{(3)}\simeq 0.382$) the eigenvalues are one positive
($\lambda_1=2.236$) and one negative ($\lambda_2=-4.236$) and this
corresponds to a saddle point with an attractive and a repulsive
direction. The separatrix in Fig.~\ref{phase_flow} (left) in the
attractive direction corresponds to the eigenvector associated to the
negative eigenvalue: $\rho_R=\rho_I$ (the red line in figure). The
other separatrix is locally approximated (in the neighborhood of
($\rho_I^{(3)}$, $\rho_R^{(3)}$), by the eigenvector associated to the
positive eigenvalue $\rho_R = - \rho_I + 2 \rho_I^{(3)}$ (the green
line in figure). The phase flow is such that if the initial condition
is such that $\rho_I < \rho_R$ ($\rho_I > \rho_R$) the system will
deterministically converge to the regular (irregular) state. On the
other hand if the initial condition is such that $\rho_I = \rho_R$ the
system will converge to the stationary solution ($\rho_I^{(3)}$,
$\rho_R^{(3)}$) (see also Refs.~\cite{baronchelli_2007,NG_depth}).

\subsection{} In this appendix, we provide explicit proofs that the
intersections $x_1$ and $x_2$ of the conic section $\cal{C}_1$ with
the axis $y=0$ [see Eq.~(\ref{x1x2})] are always physical and always
unphysical, respectively.

Let us first consider the case $a>0$. A crucial point to recognize is
that, since $a+2d=-(c+f)<0$, if $a$ is positive $d$ must be negative.
Hence $\Delta=\sqrt{d^2-af} \le |d|=-d$. As a consequence
$x_1=-(\Delta+d)/a \ge 0$. The alternative expression of
$\Delta=\sqrt{(a+d)^2+ac}$ implies, since both $a$ and $c$ are
positive, $\Delta \ge |a+d|$. This means that $\Delta \ge -(a+d)$ so
that $x_1=-(\Delta+d)/a \le 1$. It also implies $\Delta \ge (a+d)$
which, inserted into the expression $x_2=(\Delta-d)/a \ge 1$, yields
$x_2 \ge 1$.

The arguments are similar for $a<0$. In this case
$\Delta=\sqrt{d^2-af} \ge |d| \ge -d$ so that $x_1=(-d-\Delta)/a \ge
0$. By the same token $\Delta=\sqrt{d^2-af} \ge d$, implying
$x_2=(\Delta-d)/a \le 0$. Finally, to show that $x_1 \le 1$ we start
from $\Delta=\sqrt{(a+d)^2+ac} \le |a+d|$. The quantity $a+d$ is
negative for $a<0$, because $a+2d<0$ implies $2a+2d<a<0$. Hence
$\Delta \le |a+d|=-(a+d)$ so that $x_1=-(d+\Delta)/a<1$.

\bibliographystyle{apsrev4-1}
%\bibliography{3statepaper2}


\end{document}